\RequirePackage{fix-cm}
\documentclass[smallextended]{svjour3}       % onecolumn (second format)
\smartqed  % flush right qed marks, e.g. at end of proof
\usepackage[authoryear]{natbib}
\usepackage{graphicx}
\usepackage{hyperref}
\usepackage{xcolor}
\usepackage[most]{tcolorbox}

\definecolor{darkgreen}{rgb}{0.0, 0.5, 0.0}

\newcommand{\rqone}{\textit{What types of software engineering inquiries do developers present to ChatGPT in the initial prompt?}}
\newcommand{\rqtwo}{\textit{How do developers present their inquiries to ChatGPT in multi-turn conversations?}}
\newcommand{\rqthree}{\textit{What are the characteristics of the sharing behavior?}}

\newcommand{\rqmultiprompt}{What are the roles of developer prompts in multi-turn conversations?}
\newcommand{\rqmultipattern}{What are the flow patterns in multi-turn conversations?}

\newcommand{\rqpurpose}{What are the rationale behind the sharing of conversations?}
\newcommand{\rqloc}{Where are the links to these shared conversations located?}
\newcommand{\rqwho}{Who shared the conversations?}

\begin{document}

\title{An Empirical Study on Developers’ Shared Conversations with ChatGPT in GitHub Pull Requests and Issues}

\titlerunning{Analyzing Developers' Shared Conversations with ChatGPT}

\author{Huizi Hao$^*$         \and
        Kazi Amit Hasan$^*$   \and
        Hong Qin \and
        Marcos Macedo \and
        Yuan Tian \and
        Steven H. H. Ding \and
        Ahmed E. Hassan
}


\institute{Huizi Hao, Kazi Amit Hasan, Hong Qin, Marcos Macedo, and Yuan Tian\at
              School of Computing, Queen's University, ON, Canada \\
             \email{\{huizi.hao, kaziamit.hasan, 21hq5, marcos.macedo, y.tian\}@queensu.ca}
           \and
           Steven H. H. Ding \at
               % School of Information Studies, McGill University, Canada \\
               School of Computing, Queen's University \\
              \email{steven.ding@queensu.ca}
              \and
              Ahmed E. Hassan \at
Software Analysis and Intelligence Lab (SAIL), ON, Canada \\
              \email{ahmed@cs.queensu.ca}
}

\def\thefootnote{*}\footnotetext{These authors contributed equally to this work.}\def\thefootnote{\arabic{footnote}}

\date{Received: date / Accepted: date}

\maketitle

\begin{abstract}
%\marcos{Is this accurate? What does enhance contributions mean?}\yuan{change to a neutral tone, i.e., support}

ChatGPT has significantly impacted software development practices, providing substantial assistance to developers in a variety of tasks, including coding, testing, and debugging. Despite its widespread adoption, the impact of ChatGPT as an assistant in collaborative coding remains largely unexplored. In this paper, we analyze a dataset of 210 and 370 developers' shared conversations with ChatGPT in GitHub pull requests (PRs) and issues. We manually examined the content of the conversations and characterized the dynamics of the sharing behavior, i.e., understanding the rationale behind the sharing, identifying the locations where the conversations were shared, and determining the roles of the developers who shared them. Our main observations are: (1) Developers seek ChatGPT's assistance across 16 types of software engineering inquiries. In both conversations shared in PRs and issues, the most frequently encountered inquiry categories include code generation, conceptual questions, how-to guides, issue resolution, and code review. (2) Developers frequently engage with ChatGPT via multi-turn conversations where each prompt can fulfill various roles, such as unveiling initial or new tasks, iterative follow-up, and prompt refinement. Multi-turn conversations account for 33.2\% of the conversations shared in PRs and 36.9\% in issues. (3) In collaborative coding, developers leverage shared conversations with ChatGPT to facilitate their role-specific contributions, whether as authors of PRs or issues, code reviewers, or collaborators on issues. Our work serves as the first step towards understanding the dynamics between developers and ChatGPT in collaborative software development and opens up new directions for future research on the topic.

\end{abstract}

\maketitle

\section{Introduction} \label{sec:intro}
% how LLMs and ChatGPT support SE tasks.
% Large language models (LLMs) like ChatGPT are improving the productivity of software developers \marcos{cite github co-pilot stats}.   Many studies are investigating the quality of responses generated by ChatGPT, the efficacy of various prompting techniques, and its comparative performance in programming contests \marcos{add more citations}. 

%like LLMs, \marcos{This is very subtle and open to interpretation. But I feel FM are LLMs but LLMs are not necessarily FM. So we might need to remove the LLM comparison}
Recent advances in Foundation Models (FMs) hold considerable promise for automating various software engineering tasks. FM-powered tools like GitHub Copilot~\footnote{\url{https://github.com/features/copilot}}, Amazon CodeWhisperer~\footnote{\url{https://aws.amazon.com/codewhisperer}}, and OpenAI ChatGPT~\footnote{\url{https://openai.com/chatgpt}}, are now embraced by many professional software practitioners~\citep{zhang2023practices}. Such technologies essentially acquire important capabilities based on massive, typically natural-language, data sets and are able to suggest recommendations to software developers, providing source code completion, automatic generation of documentation, or other types of software engineering support. 

% how existing research papers analyze LLM, especially the usage of LLM in practice
%The rapid advancements in artificial intelligence, specifically language models like ChatGPT, have profoundly impacted various fields, including software engineering. Recent studies have begun to shed light on the influence of ChatGPT on software development, particularly focusing on its integration into the workflow of developers. 
%However, they also highlight challenges in understanding, editing, and debugging the code snippets generated by Copilot, which can impede task-solving effectiveness.

Current research on FMs for software engineering has primarily focused on evaluating the effectiveness of these models, including specialized versions modified through prompt engineering or fine-tuning, against traditional automated software engineering solutions using standard benchmarks~\citep{jiang2023impact,lu2023llama,hou2023large,siddiq2023exploring,deng2024large,guo2024exploring}. In contrast, a limited number of studies~\citep{vaithilingam2022expectation, ziegler2022productivity,mozannar2022reading,barke2023grounded,liang2024large} have investigated how FM-powered tools are practically employed by software developers within the software development life cycle. Most of these studies focus on GitHub Copilot, an FM-powered code generation tool via small-scale (20-42 participants) user studies or surveys.  %Their studies reveal that programmers prefer using Copilot for daily programming tasks, as it provides useful starting points and reduces the need for online searches. Nevertheless, they are limited in scope, focusing solely on Copilot and involving small-scale (20 - 24 participants) user studies. 

The usage of ChatGPT in software development encompasses a broader spectrum. Unlike Copilot, which primarily aids in code completion and generation, ChatGPT is designed to generate human-like text based on the received input, i.e., \textit{prompt}. This makes ChatGPT a more versatile tool that is applicable to a wider range of tasks beyond coding. Despite its potential, there is no research on the dynamics of developer interactions with ChatGPT, the challenges faced, and the opportunities it offers for open-source projects. Moreover, given that modern software systems are crafted by teams rather than isolated individuals, the capacity of ChatGPT to bolster and transform collaborative practices has not been explored. This potential extends well beyond its current use in individual tasks, suggesting a significant yet unexplored impact of FM-powered tools on team-based software development. 
%Understanding these aspects is essential to fully harness ChatGPT and other FM-powered tools' capabilities in software engineering. 
%\marcos{For leveraging potential in assisting developers who create issues and pull requests?}\yuan{for all developers, including those use it in collaborative coding}

% move on to the conversation-sharing function of ChatGPT
%The sharing  the share-link functionality that allows users to share anonymous links that can be seen by other people. This enables to share chats and understand the dialog of the sharing user. In the context of GitHub Pull Requests and Issues, it has been observed by \marcos{add citation here to DevGPT maybe?} that users are increasingly sharing their conversations. We perform an empirical study and try to understand what is the information that the users are sharing on those conversations and their motivation for doing so.  
%Analyzing these conversations can provide insights into the decision-making processes, problem-solving strategies, and collaborative practices employed by developers while utilizing ChatGPT. It can shed light on the role of LLMs in facilitating more efficient and effective communication within software development team, enhancing code quality, and streamlining project management. 

\begin{figure}[ht]
    \centering
    \includegraphics[width=\linewidth]{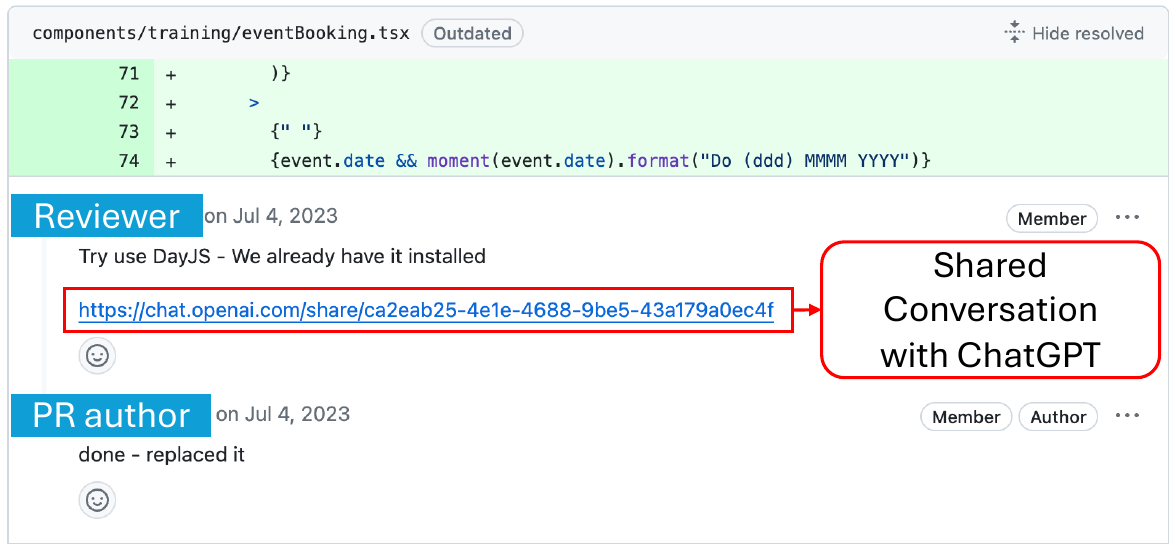}
    \caption{An example of shared conversations with ChatGPT in a GitHub pull request. In the shared conversation, the reviewer asked ``Can i replicate this functionality with DayJS? moment(event.date).format(``Do (ddd) MMMM YYYY'')''.}
    \label{fig:example}
    \vspace{-5mm}
\end{figure}

To fill the gap, we analyze developers' shared ChatGPT conversations within GitHub issues and pull requests (PRs). We postulate that these shared conversations will not only reveal how developers engage with ChatGPT in the quality assurance of open-source software projects, such as resolving issues, but also uncover the usage of the shared conversation in collaborative coding, such as contributing to PRs and issues. Link sharing was introduced in ChatGPT in May 2023~\citep{openaihelp}, allowing users to generate a unique URL for a ChatGPT conversation. This feature enables users to share a snapshot of an entire conversation up to the point of the link being shared. Such shared conversations can help highlight important messages, reference discussions for collaborative purposes, or create a point for future reference~\citep{sharedlinkfaq}. Figure~\ref{fig:example} shows an example of a conversation with ChatGPT being shared within a code review comment on a GitHub pull request. The shared conversation contains one \textit{turn} - a complete cycle of a developer posing a question or problem and ChatGPT responding. The code reviewer proposed a request asking the pull request author to use a JavaScript library named \textit{Day.js} to parse and format dates. To demonstrate an example of how Day.js can be utilized, the reviewer asked ChatGPT how to replace a code snippet using Day.js, and ChatGPT responded. Then, the pull request author implemented the suggested code provided by ChatGPT. This example demonstrates the potential benefit of sharing a conversation with ChatGPT in a collaborative coding environment like GitHub. 
\vspace{-0.5cm}

%By analyzing such instances, we can gain a deeper understanding of how shared conversations with LLM-based tools like ChatGPT are being utilized to enhance collaborative efforts, improve code quality, and streamline the software development process. 

% what we did and what we find
%In this work, we take the first step in understanding developer-ChatGPT conversations shared by developers in GitHub Issues and Pull Requests. Insights about issue types developers bring to ChatGPT, shared conversation usages by developers in GitHub Issues and Pull Requests and developer presenting styles in multi-turn shared conversations are critical for software engineering professionals and researchers, as well as tool designers. With the aim of identifying these information, we followed an empirical study to examine developer prompts in the shared conversations, and developer comments in GitHub Issues and Pull Requests where the shared conversations were posted. We conducted in-depth data analysis on the resulting labels from the empirical study to gain insights into developer shared conversations with ChatGPT in GitHub Issues and Pull Requests. Our study addresses the following research questions:

%\subsection{Study Design \marcos{I think we should put this into it's own section}}\yuan{common practice is to keep it as part of the introduction, no subsection in intro} 
To better understand the characteristics of shared conversations and their implications on collaborative software engineering practice, we manually analyzed a dataset containing 580 shared ChatGPT conversations within GitHub PRs and issues, specifically, 210 in PRs and 370 in issues. Our empirical study addresses the following three research questions (RQs):

\begin{itemize}
    \item [] \textbf{RQ1: \rqone} In this RQ, we manually examined the content of all initial prompts, 580 prompts in total, in the collected shared conversations. We developed a taxonomy composed of 16 types of software engineering-related inquiries. The most frequently encountered inquiries are about code generation, addressing conceptual questions, providing how-to guidance, assisting with issue resolution, and conducting code reviews. 
    \vspace{0.1cm}
    \item [] \textbf{RQ2: \rqtwo} In this RQ, we manually examined how developers interact with ChatGPT in \textit{multi-turn conversations}, which consist of several rounds of prompts and responses between a developer and ChatGPT. Specifically, we investigated how developers structure their follow-up prompts subsequent to the initial prompt. We developed a taxonomy composed of seven types of roles a prompt plays within a multi-turn conversation. Our findings reveal that developers actively engage in multiple rounds of interactions with the aim of improving the quality of ChatGPT's responses. This is primarily achieved through the posting of follow-up questions and the refinement of prior prompts.

    %In this RQ, we are investigating the motivations driving developers to share their interactions with ChatGPT in the context of GitHub issues and pull requests. We classified the developer usage of shared conversation from GitHub Pull Request and GitHub Issues into nine categories.
    \vspace{0.1cm}
    \item [] \textbf{RQ3: \rqthree} In this RQ, we investigate the patterns in the sharing behaviors of developers. More specifically, we examine where those conversations are shared by whom and for what purpose. Our analysis reveals that developers utilize shared conversations as a way to complement their role-specific contributions, facilitating a more efficient and transparent collaborative process.  
\end{itemize}

%Our results suggest future researchers \yuan{something from implications should be put here.}

The main contributions of this paper are described as follows: 
\begin{itemize}
    \item [-] We present a comprehensive analysis of how developers share their conversations with ChatGPT within the context of open-source projects, particularly focusing on GitHub pull requests and issues. Our study introduces two taxonomies to classify the dynamics of these interactions: one for the types of software engineering inquiries in developers' prompts and another for the roles of prompts in multi-turn conversations with ChatGPT. To support further research, our replication package contains 580 and 654 manually annotated prompts aligned with these taxonomies.

   \item [-] Our study reveals the multifaceted usage of ChatGPT in software engineering. Beyond assisting with technical tasks, we observe the usage of shared conversations to support collaboration among developers in open-source projects. 

    \item [-] We provide implications for developers and software engineering researchers based on our findings. These implications offer insights to improve the use of FM-powered tools like ChatGPT further in collaborative software development.

\end{itemize}

%To the best of our knowledge, this is the first research work investigating the sharing of developers' conversations in the open-source software development process. By gaining insights into the what, how, and why developers share their conversations with ChatGPT in GitHub issues and pull requests, we shed light on future directions in enhancing LLM-driven software development and collaboration in software projects. Moreover, we have made our replication package publicly available at Zenodo, encouraging and supporting further research on this topic.

We organize the remainder of the paper as follows.  Section~\ref{sec:empirical} introduces our dataset. Section~\ref{sec:rq1}-\ref{sec:rq3} present methodology and answers to each of the three research questions. Section~\ref{sec:disc} discusses the implications of findings for practitioners and researchers. Section~\ref{sec:threats} presents threats to validity, and Section~\ref{sec:related} presents related work. Finally, Section~\ref{sec:conclusion} concludes the paper.

\section{Data Collection}\label{sec:empirical}
\begin{figure*}
    \centering
    \includegraphics[width=0.88\linewidth]{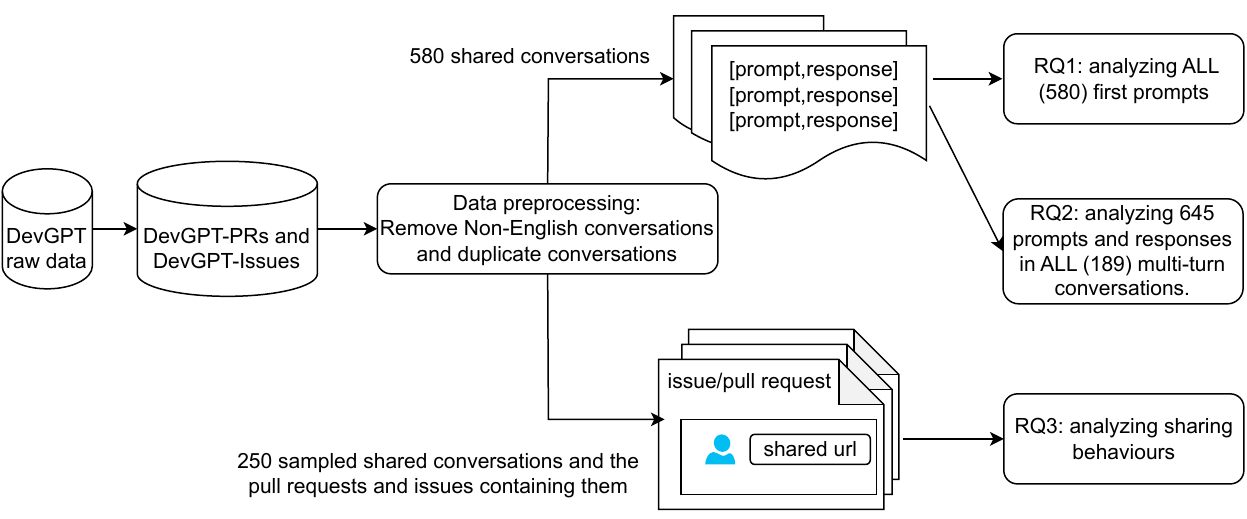}
    \caption{Overview of the data collection process for answering three research questions.}
    \label{fig:Data preparation}
\end{figure*}

In this section, we introduce the used dataset for our study (Section~\ref{sec:dataset}), followed by the used method utilized in preprocessing dataset (Section~\ref{sec:datapreprocess}) and preparing datasets for research questions (Section~\ref{sec:datarqs}). Figure~\ref{fig:Data preparation} shows an overview of the data collection process and the used data for each of our RQs.

\subsection{Data Source}\label{sec:dataset}
Our research leverages the DevGPT dataset~\citep{devgpt} as the primary data source. DevGPT constitutes of an extensive archive of Developer-ChatGPT interactions, featuring 16,129 prompts and ChatGPT's replies. Each shared conversation is coupled with the corresponding software development artifacts to enable the analysis of the context and implications of these developer interactions with ChatGPT. This collection was assembled by extracting shared ChatGPT links found in various GitHub components, such as source code, commits, pull requests, issues, discussions, and threads on Hacker News, over the period from July 27, 2023, to October 12, 2023. The DevGPT dataset is publicly available in a GitHub repository~\footnote{https://github.com/NAIST-SE/DevGPT}, offering several snapshots. In this study, we focus on the most recent snapshot available as of October 12, 2023.~\footnote{The dataset is cumulative, meaning that the latest snapshot includes all conversations shared in prior snapshots.}

%Figure \ref{fig:example} represents the overall scenario how developers share their ChatGPT conversation of a GitHub pull request \footnote{https://github.com/polywrap/evo.ninja/pull/206}. In this pull request, the author opened the pull request, which was subsequently reviewed by two individuals. Reviewer 1 contributed by including comments that contained links to two separate ChatGPT conversations while addressing issues within the PR. Following this input, the PR received approval from both reviewers and was successfully merged. This shared conversation not only reflects the developer's intentions but also aligns with the overall objectives of the pull request.

\subsection{Data Preprocessing}\label{sec:datapreprocess}
As our analysis exclusively focuses on shared conversations occurring within GitHub issues and pull requests, we only consider the corresponding data provided by DevGPT, referred to as DevGPT-PRs and DevGPT-Issues. Based on our observations, we then perform the following two data preprocessing steps:

\begin{enumerate}
    \item The shared conversations contain sentences (prompts and replies) written in various human languages. To avoid potential misunderstanding from translating other languages different than English, we decided to only keep the conversations written in English. We utilized a Python library named \textit{lingua}\footnote{https://github.com/pemistahl/lingua-py} to identify conversations containing non-English content and removed those conversations. Specifically, we excluded 46 non-English conversations from DevGPT-PRs and 114 from DevGPT-Issues.
    
    %We then manually read the remaining conversations and removed cases that the library did not detect successfully. Specifically, we excluded 46 non-English conversations from DevGPT-PRs and 114 from DevGPT-Issues.

    \item The shared conversations contain duplicates, i.e., conversations with identical prompts and responses. We detected duplicate conversations and kept only one instance for analysis. Specifically, we removed 20 duplicated conversations from DevGPT-PRs and 83 duplicated conversations from DevGPT-Issues.
\end{enumerate}

After performing the above two data preprocessing steps, we ended up with 220 conversations from DevGPT-PRs, and 401 conversations from DevGPT-Issues.

\subsection{Preparing Data for RQs}\label{sec:datarqs}
\begin{figure}
    \centering
    \includegraphics[width=0.8\linewidth]{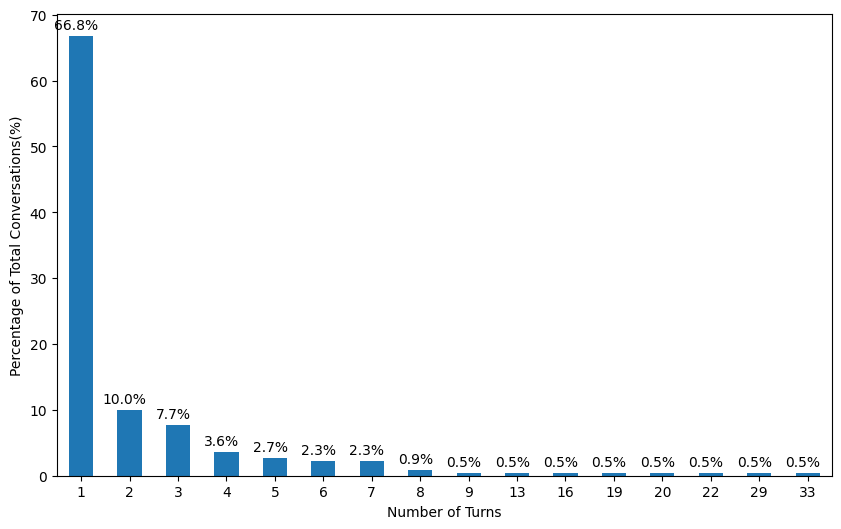}
    \caption{Number of turns in shared conversations within DevGPT-PRs.}
    \label{fig:Distribution of Prompts - PR}
\end{figure}

\begin{figure}
    \centering
    \includegraphics[width=0.8\linewidth]{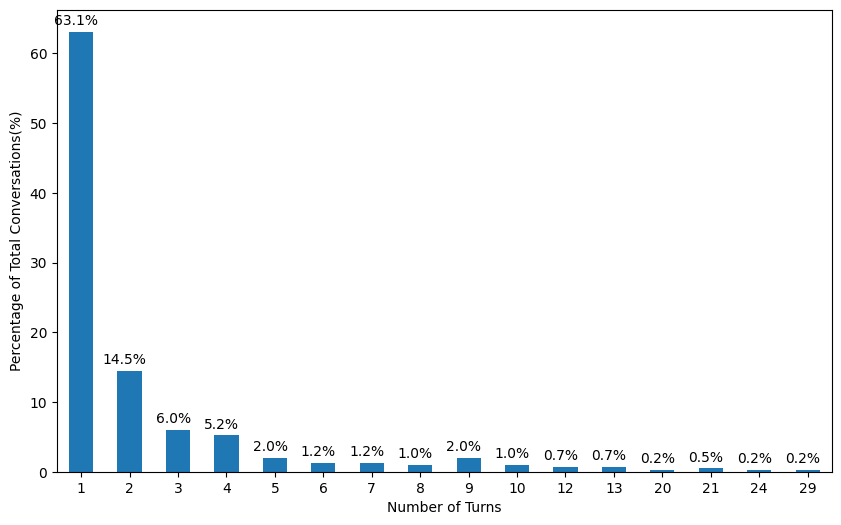}
    \caption{Number of turns in shared conversations within DevGPT-Issues.}
    \label{fig:Distribution of Prompts - Issues}
\end{figure}

Figures~\ref{fig:Distribution of Prompts - PR} and~\ref{fig:Distribution of Prompts - Issues} show the distribution of conversational turns within the preprocessed datasets. As shown in these figures, a large majority of shared conversations in both DevGPT-PRs (66.8\%) and DevGPT-Issues (63.1\%) are single-turn interactions. Meanwhile, conversations extending beyond 8 turns - comprising 8 prompts and their 8 corresponding replies - are notably infrequent, accounting for only 4\% in DevGPT-PRs and 6\% in DevGPT-Issues. Given this distribution, we choose to implement a cutoff at 8 turns for RQ1-3 involving both datasets. This approach allows us to concentrate our investigation on the most prevalent patterns of interaction, thereby ensuring that our analysis remains closely aligned with the conversational dynamics that characterize the vast majority of the dataset. Following this cutoff, the finalized datasets encompass a total of 212 conversations for DevGPT-PRs and 375 for DevGPT-Issues 


%\marcos{amit:: Yuan, can you please review this sentence?} 
As shown in Figure~\ref{fig:Data preparation}, in RQ1, we analyzed the contents of 580 initial prompts~\footnote{This number is lower than the 587 total conversations (212 from DevGPT-PRs and 375 from DevGPT-issues) due to the exclusion of seven non-English conversations that were not identified during the data preprocessing phase by the tool utilized.}. In RQ2, we analyzed the content in the 645 prompts within all 189 multi-turn conversations. For RQ3, we extend our manual analysis to pull requests and issues that contain shared conversations. Specifically, we randomly sampled a statistically significant set containing 90 GitHub pull request comments and 160 GitHub issue comments containing shared conversations from DevGPT-PRs and DevGPT-Issues. The sampling is based on the results of RQ2. The detailed process is presented in the approach of RQ3 in Section~\ref{sec:rq3}. 

\section{RQ1: \rqone}\label{sec:rq1}

\textbf{Motivation:} The objective of RQ1 is to characterize types of software engineering-related inquiries that are presented by developers in their shared conversations in GitHub issues and pull requests, particularly focusing on the initial prompts of each conversation. This insight about the presented inquiries in shared conversations is crucial for several reasons. Firstly, the taxonomy provides a structured framework to categorize and understand the diverse range of software engineering inquiries developers raise to ChatGPT. Secondly, by analyzing the frequency of these inquiries, we can uncover common needs among developers when utilizing ChatGPT in collaborative software development. This insight can inform the future improvement of FM-powered tools, including ChatGPT, and the establishment of benchmarks for evaluating the effectiveness of such tools in collaborative software development. Finally, comparing inquiries across pull requests and issues allows for a comprehensive examination of the different contexts in which these inquiries are encountered. 

%\yuan{need to revisit this number, some conversations have initial task in the second not initial prompt.}
We chose to focus on the initial prompt of each conversation in RQ1 due to two main concerns: 1) The majority of the conversations (66.8\% of DevGPT-PRs and 63.1\% of DevGPT-Issues) consist of single-turn interactions, wherein the initial prompt encompasses the sole task presented to ChatGPT. 2) Upon examination of multi-turn conversations, we observed that subsequent prompts never introduce new types of software engineering inquiries beyond those presented in the initial prompts.

%(19\%, from DevGPT-PRs, and 10\%, from DevGPT-Issues, out of all multi-turn conversations\yuan{need a discussion on this statement!}). 

%3) Unlike the initial prompts, which are often isolated and self-contained, follow-up prompts in multi-turn conversations tend to build upon previous exchanges and are primarily used to guide ChatGPT towards refining its response to the initial task (ref. Section~\ref{sec:rq2}). Therefore, incorporating the types of inquiries presented in follow-up prompts may 

%Incorporating the presented tasks in follow-up prompts may potentially impact the frequency distribution of task types, potentially resulting in results that are challenging to interpret \marcos{I didn't quite understand this. Should we keep or remove?}.

%https://queensu.zoom.us/j/93382805661?pwd=cC9VNXdFNkM4b2lWTGJLcW54WFQ0Zz09
\subsection{Approach}
We use open coding to manually categorize the initial prompts of all conversations in DevGPT-PRs and DevGPT-Issues. During labeling, we further removed non-English prompts that were not detected by data preprocessing. Specifically, 2 and 5 prompts from DevGPT-PRs and DevGPT-Issues were removed, respectively. Thus, our taxonomy was developed based on 210 prompts from DevGPT-PRs and 370 prompts from DevGPT-Issues. Our labeling was conducted over three rounds:
\begin{itemize}
    \item In the first round, 50 randomly selected prompts from each dataset (DevGPT-PRs and DevGPT-Issues) are selected and categorized by five co-authors independently, i.e., 100 prompts in total. After a comprehensive discussion, the team formulated a coding book with 16 labels, excluding \textit{others}. 

    \item In the second round, another set of 100 prompts was individually annotated by two co-authors based on the coding book established in the first round, with an equal split sourced from both DevGPT-PRs and DevGPT-Issues. The two annotators achieved an inter-rater agreement score of 0.77, as measured by Cohen's kappa coefficient, representing a substantial agreement between the two annotators~\citep{landis1977measurement}. They then discussed to resolve disagreement and further refined the coding book.

    \item In the final round, the two annotators who participated in the second round continued and independently labeled the remaining prompts based on the coding book.
\end{itemize}

\subsection{Results}
Table~\ref{tab:Developer_Tasks} presents the derived taxonomy from analyzing the 580 initial prompts in shared conversations across the two datasets. The results show a significant focus in the initial prompts on specific software engineering (SE) inquiries, with 198 of the initial prompts in DevGPT-PRs and 329 in DevGPT-Issues presenting such inquiries. The remainder, labeled as \textit{Others}, comprises 12 prompts in DevGPT-PRs and 41 prompts in DevGPT-Issues, featuring either ambiguous content or information not directly related to SE inquiries, for example, ``\textit{Explain Advancing Research Communication – the role of Humanities in the Digital Era}''.

\begin{table}[h]
    \centering
    \caption{Taxonomy of the initial prompt within the shared conversations.} \label{tab:Developer_Tasks}
    \begin{tabular}{|l|c|c|c|}
    \hline
       \textbf{Category}  & \textbf{PR} & \textbf{Issue}  \\
    \hline
      SE-related:   &  198(100\%) &  329(100\%)  \\
      (C1) Code generation   & 40 (20\%) & 90 (27\%) \\
      (C2) Conceptual   & 37 (18\%) & 45 (14\%)  \\
      (C3) How-to   & 26 (13\%) & 74 (22\%)  \\
      (C4) Issue resolving   & 24 (12\%) & 46 (14\%) \\
      (C5) Review   & 18 (9\%) & 12 (4\%)\\
      (C6) Comprehension   & 13 (7\%) & 10 (3\%)\\
      (C7) Human language translation   & 11 (6\%) & 0 (0\%)\\
      (C8) Documentation   & 10 (5\%) & 7 (2\%)\\
      (C9) Information giving   & 8 (4\%) & 8 (2\%)\\
      (C10) Data generation   & 4 (2\%) & 12 (4\%)\\
      (C11) Data formatting   & 2 (1\%) & 9 (3\%)\\
      (C12) Math problem solving   & 2 (1\%) & 9 (3\%)\\
      (C13) Verifying capability   & 2 (1\%) & 1 (0\%)\\
      (C14) Prompt engineering   & 1 (0\%) & 0 (0\%)\\
      (C15) Execution & 0 (0\%) & 4 (1\%)\\
      (C16) Data analysis   & 0 (0\%) & 2 (1\%)\\ \hline
      Others & 12 & 41 \\
    \hline
    \end{tabular}
\end{table}

In both datasets, the five most prevalent SE-related inquiries identified were \textit{Code Generation}, \textit{Conceptual}, \textit{How-to}, \textit{Issue Resolving}, and \textit{Review}, accounting for 72\% and 81\% of SE-related prompts in DevGPT-PRs and DevGPT-Issues, respectively. A notable distinction between the two sources is the higher occurrence of prompts in DevGPT-PRs seeking help with \textit{Review}, \textit{Comprehension}, \textit{Human language translation} and \textit{Documentation}. Conversely, DevGPT-Issues tend to feature more requests related to \textit{Data generation}, \textit{Data formatting}, and \textit{Mathematical problem resolving}. 

Below, we describe each SE-related category in more detail.

\noindent \textbf{(C1) Code Generation:} This category shows the highest number of prompts in both resources, accounting for 20\% in DevGPT-PRs and 27\% in DevGPT-Issues, respectively. In prompts labeled as this category, developers request ChatGPT to generate code snippets based on a provided description. A follow-up analysis of these requests revealed five distinct types of requirements for code generation. As shown in Table~\ref{tab:codegen_subtype}, the majority of requests (48\%) involve developers specifying their needs in textual form. However, it's common for developers to supply existing code, either as context for generation or as a basis for adaptation. The requirements range from code generation to more specific inquiries, such as code transformation across different programming languages or platforms and generating test code.

\begin{table}
\centering
\caption{Five types of code generation inquiries identified from initial prompts.} \label{tab:codegen_subtype}
\begin{tabular}{|p{9cm}|c|c|}
\hline
\textbf{Type} & \textbf{Freq.} \\ \hline
Code generation based on textual specifications. & 48\% \\
Code generation with modifications to an existing implementation. & 24\% \\
Code generation incorporating code context. & 12\% \\
Code transformation across languages or platforms. & 11\% \\
Test code generation. & 5\% \\
\hline
\end{tabular}
\end{table}

\noindent \textbf{(C2) Conceptual:} In 18\% of the initial prompts in DevGPT-PRs and 14\% in DevGPT-Issues, developers engage ChatGPT to seek knowledge and clarification about theoretical concepts, principles, practices, tools, or high-level implementation ideas. This exploration spans questions on technical specifics, such as the memory usage limits of WebAssembly (WASM) in Chrome, to practical implementation queries, like the feasibility of developing a Redis-like cache using SQLite with time-to-live (TTL) functionality. These examples underscore ChatGPT's role not only as a direct coding assistant but also as a conceptual guide aiding developers in understanding and applying complex technical knowledge with which they are not familiar. 

\noindent \textbf{(C3) How-to:} We observe that 13\% of initial prompts in DevGPT-PRs and a higher percentage, 22\%, in DevGPT-Issues, are from developers seeking step-by-step instructions to accomplish specific SE-related goals. For instance, ``\textit{I am doing... How can I achieve this?'', and ``How to make an iOS framework M1 compatible?}''. Such examples illustrate ChatGPT's role in offering insights and starting points for tackling SE challenges. The higher occurrence of ``How-to'' prompts in DevGPT-Issues compared to DevGPT-PRs suggests that developers might turn to ChatGPT more frequently when they are in the initial stages of problem-solving, possibly lacking a concrete solution. 

\noindent \textbf{(C4) Issue resolving:} In 12\% of the initial prompts in DevGPT-PRs and 14\% in DevGPT-Issues, developers seek assistance to resolve SE-related issues. Further examination of these prompts identified five types of information developers provide when seeking help with issues, as detailed in Table~\ref{tab:issue_subtype}. The distribution of information types shows developers' diverse strategies to present their challenges in the prompt. The most popular method (36\%), sharing error messages or traces, indicates developers' preference for direct feedback on specific errors. Combining code snippets with a detailed description of the error or unexpected behavior (19\%) is more common than merely verbally describing the issues (16\%). Another common type (17\%) is prompts seeking assistance for troubleshooting tasks unrelated to code development, such as installing a library or setting up a system. Interestingly, developers occasionally (13\%) include external links in their initial prompts, expecting ChatGPT to access these links, review the associated issues, and offer suggestions for resolution.

\begin{table}[]
\centering
\caption{Five types of information in initial prompts for issue resolving assistance.} \label{tab:issue_subtype}
\begin{tabular}{|p{9cm}|c|}
\hline
\textbf{Type} & \textbf{Freq.} \\ \hline
Error messages, stack traces, or warnings. & 36\% \\
Code snippet with error messages or unexpected behavior description. & 19\% \\
Troubleshooting task. & 17\% \\
Description of unexpected behavior. & 16\% \\
Issue reports. & 13\% \\
\hline
\end{tabular}
\end{table}

\noindent \textbf{(C5) Review:} In DevGPT-PRs, 9\% of the initial prompts involve requests for ChatGPT to offer suggestions for code improvement or to compare different implementations or design choices. This frequency is notably higher than in DevGPT-Issues, where 4\% of the initial prompts reflect similar requests. This distribution is expected, as developers tend to prioritize code quality and seek external validation or suggestions for enhancement in the context of PRs. Conversely, in issues, the primary focus is on identifying, discussing, and exploring potential solutions to existing problems. 

\noindent \textbf{(C6) Comprehension:} In DevGPT-PRs, 7\% of the initial prompts seek ChatGPT's assistance in understanding the behavior of code snippets or software artifacts, exceeding the 3\% observed in DevGPT-Issues. These prompts typically begin with phrases like ``\textit{Explain this code}''. The higher prevalence in PRs suggests a focused interest in clarifying code behavior as part of the review and integration process.

\noindent \textbf{(C7) Human language translation:} Notably, ChatGPT's capability in human language translation, while not specific to SE, is utilized by developers in 6\% of the conversations in DevGPT-PRs, but not in DevGPT-Issues. This category encompasses requests for translating text relevant to software development from one language to another. Examples include UI components, code strings, technical documentation, and project descriptions.

\noindent \textbf{(C8) Documentation:} Within DevGPT-PRs and DevGPT-Issues, 5\% and 2\% of the initial prompts from each dataset request ChatGPT for assistance with creating, reviewing, enhancing, or refining technical documentation. This category spans a wide array of documentation forms and encompasses project descriptions, issue reports, the content of pull requests, technical communications, and markdown files, among others. The requests indicate developers' recognition of ChatGPT's capability to contribute to technical documentation's clarity, accuracy, and effectiveness, reinforcing the importance of well-crafted documentation in software development.

\noindent \textbf{(C9) Information giving:} In DevGPT-PRs and DevGPT-Issues, 4\% and 2\% of initial prompts, respectively, diverge from directly requesting assistance with a specific inquiry. Instead, these prompts offer contextual information relevant to an SE-related inquiry presented in subsequent prompts. This contextual information can be a sharing of code snippets or an explanation of the developer's current project or the intended role of the ChatGPT. This approach, found in multi-turn conversations, unveils a different prompting strategy developers employ - putting background information as the initial and isolated prompt before presenting the inquiry. 

\noindent \textbf{(C10) Data generation:} In 2\% and 4\% of the initial prompts in DevGPT-PRs and DevGPT-Issues, developers ask ChatGPT to create various types of data, including test inputs, search queries, icons, example of API specification, and names for websites. 

\noindent \textbf{(C11) Data formatting:} Besides using ChatGPT to generate data, in 1\% and 3\% of initial prompts in DevGPT-PRs and DevGPT-Issues, developers ask ChatGPT to transform a given data into a different format. For instance, ``\textit{You are a service that translates user requests into JSON objects of type "Plan" according to the following TypeScript definitions...}''. 

\noindent \textbf{(C12) Mathematical problem solving:} Developers turn to ChatGPT for assistance with mathematical problems in 1\% of the initial prompts in DevGPT-PRs and 3\% in DevGPT-Issues. These inquiries often aim to deepen one's understanding of algorithms that underpin programming tasks or to create test cases for software validation. 

\noindent \textbf{(C13) Verifying capability (of ChatGPT):} We encountered two and one initial prompts from DevGPT-PRs and DevGPT-Issues where one of the developers ask ``\textit{are you familiar with typedb?}''. In this case, the developer assessed ChatGPT's capabilities to determine the scope of assistance it can provide for their SE-related inquiries - an establishment of trust and reliability in FM-powered tools for software development.

\noindent \textbf{(C14) Prompt engineering:} We find one initial prompt in DevGPT-PRs where the developer presents an initial prompt for a SE-related inquire and asks ChatGPT to provide a better prompt. This instance presents a unique prompt engineering strategy, using ChatGPT as a tool to refine the art of prompting itself. 

\noindent \textbf{{(C15) Execution:}} Exclusively in DevGPT-Issues, 4 of the initial prompts request ChatGPT to execute a specific task, which includes running an experiment or executing code. Examples of such requests are, ``\textit{Try running that against this function and show me the result}'' and ``\textit{Benchmark that for me and plot a chart}''. These prompts indicate that developers believe that ChatGPT is capable of undertaking practical execution tasks, which is rarely examined in the literature. 

\noindent \textbf{(C16) Data analysis:} Beside execution inquiries, developers also ask ChatGPT to perform data analysis tasks by providing a CSV file. We find two such cases in DevGPT-Issues. Such queries may be used to plan data preprocessing or data science projects. For instance, ``\textit{find all the entries that are present in the left and in the right column.}''.

%However, we do find cases where the developers ask ChatGPT to generate testing data, improve documentation, perform data analysis, and transfer human language (primarily for localization purposes).

%In DevGPT-PRs dataset, Code Generation (21.4\%), Conceptual (18.1\%), Issue Resolving Guidance (11.4\%) and How-to (10.0\%) are the most common intentions from developers. In the DevGPT-Issues dataset, Code Generation (25.9\%), How-to (18.6\%), Issue Resolving Guidance (12.4\%) and Conceptual (11.9\%) are the most common intentions. 

%\textbf{Overall, Code Generation (24.3\%), How-to (15.5\%), Conceptual (14.1\%) and Issue Resolving Guidance (12.1\%) are the most common intentions from developer initial prompts.} 

%The current code generation benchmark, i.e., HumanEval, employs only text specification as input. However, our observation reveals that developers often provide textual descriptions and code draft as input for code generation tasks. Based on our observation, the current benchmark can be improved to reflect the real-life situation. 

%In the initial prompt, most (36\%) of the prompts asking help for issue resolving provides error messages or execution traces without any source code. This represents a new type of task other than the popular program repair benchmarking setup for LLMs (giving buggy code and errors, fixing the bug). Interesting, these are also a few cases where the developers directly put a link to an issue report asking for guidance to resolve the issue.  

\begin{tcolorbox}[enhanced,width=4.7in,size=fbox,drop shadow southwest,sharp corners]

\textit{RQ1 Summary: A broad spectrum of software engineering-related inquiries, encompassing 16 distinct types, are presented to ChatGPT in the initial prompts of shared conversations in GitHub pull requests and issues. This diversity underscores ChatGPT's extensive potential in facilitating collaborative coding by matching a wide array of development needs. The inquiries mainly focus on code generation, conceptual understanding, procedural guidance (``how-to''), issue resolution, and code review. 
} 

\end{tcolorbox}

\section{RQ2: \rqtwo}\label{sec:rq2}
%The motivation behind exploring how developers present their issues in multi-turn conversations lies in understanding the evolving nature of their queries and requirements over the course of extended interactions with ChatGPT. Multi-turn conversations, unlike single-turns, offer a way for developers to refine, clarify, and expand upon their initial inquiries. This iterative process is crucial in situations where the complexity of a task unfolds progressively or where the initial query did not reveal all the aspects of the problem at hand.

%In the area of software development and programming, where details are very important and tasks can be complex, looking at how developers' prompts change over several exchanges can give us important understanding. This includes how developers make their questions clearer, use the feedback or answers they got from earlier conversations, and change what they're focusing on when they learn new things. Getting to know these patterns can really help make ChatGPT better at meeting the specific and changing needs of developers. This can lead to more focused and effective conversations that solve problems. 

\textbf{Motivation:} Results presented in Figure~\ref{fig:Distribution of Prompts - PR} and Figure~\ref{fig:Distribution of Prompts - Issues} reveal that a substantial portion in DevGPT-PRs (33.2\%) and DevGPT-Issues (26.9\%) encompass multi-turn conversations. In single-turn conversations, developers pose an SE-related inquiry in the initial prompt and receive one response from ChatGPT, providing a clear and direct exchange. The dynamics of multi-turn conversations, however, introduce complexity. These interactions extend beyond a simple query and response, involving a series of exchanges that potentially refine, expand, or clarify the initial inquiry. This layered communication raises a question about developers' strategies to articulate their inquiries across multiple turns. Thus, we introduce RQ2, which studies the nature of developers' prompts in multi-turn conversations. To facilitate a comprehensive analysis, we further introduce two sub-RQs:

%  This question explores the underlying patterns of developers' multi-turn prompts. We seek to uncover common strategies or approaches adopted by developers in their interactions with ChatGPT.

\begin{itemize}
    \item \textbf{RQ2.1: \rqmultiprompt} This question aims to categorize the structural role of each prompt in the corresponding multi-turn conversation. 
    \item \textbf{RQ2.2: \rqmultipattern} Based on the taxonomy proposed as the answer to RQ2.1, this question explores the frequent transition pattern of those identified roles of prompts in multi-turn conversations.

\end{itemize}

The answers to the above sub-RQs will provide insights for researchers about the dynamics and practices of developers in utilizing ChatGPT via multiple rounds of interactions. 

\subsection{Approach}

%During labeling, we further removed non-English prompts. Specifically, 2 and 0 prompts from DevGPT-PRs and DevGPT-Issues were removed, respectively. \marcos{amit ::  Yuan can you rewrite this?}  
In RQ2.1, we consider prompts in all 189 multi-turn conversations, i.e., 64 conversations from DevGPT-PRs and 125 from DevGPT-Issues. Following a method similar to RQ1, we used open coding to manually label 645 prompts (236 prompts from DevGPT-PRs and 409 prompts from DevGPT-Issues) in multi-turn conversations over three rounds:

\begin{itemize}
    \item In the first round, five co-authors independently labeled randomly selected 20 conversations from both the multi-turn DevGPT-PRs and DevGPT-Issues datasets, encompassing 40 conversations and 123 prompts. Post-discussion, we developed a coding book consisting of seven distinct labels. 

    \item In the second round, based on the existing coding book, two annotators independently labeled another set of 20 conversations from each of the multi-turn DevGPT-PRs and DevGPT-Issues datasets, a total of 144 prompts. The two annotators achieved an inter-rater agreement score of 0.89, as measured by Cohen's kappa coefficient, representing almost perfect agreement~\citep{landis1977measurement}. The annotators then discussed and refined the coding book.

    \item  Finally, each of the two annotators from round two independently labeled the remaining data.

\end{itemize}

In RQ2.2, we use a Markov model~\citep{gagniuc2017markov} to analyze the conversation flow patterns by plotting a \textit{Markov Transition Graph}. A Markov Transition Graph is a directed graph that demonstrates the probabilistic transitions between various states or nodes. In our case, each node in the graph represents a specific category developed in RQ2.1, and the directed edges between nodes denote the likelihood of transitioning from one taxonomy to another based on the multi-turn conversations we collected. 

To extract meaningful insights from the Markov Transition Graph, we propose the following post-processing steps:
\begin{enumerate}
    \item We pruned the graph by removing transitions with probabilities lower than 0.05, ensuring a focus on statistically significant relationships. 
    \item We refined the graph structure by removing nodes without incoming and outgoing edges, except the start and end nodes. This step ensures simplification as we only keep essential components.
    \item We systematically reorganized the Markov Transition Graph into a flow chart to enhance its interpretability, offering an easier-to-understand representation of the flow patterns.
\end{enumerate}

\subsection{Results}

\subsubsection{RQ 2.1 \rqmultiprompt} \par
Table~\ref{tab:Multi-turn_Prompts} presents our proposed taxonomy to classify prompts within multi-turn conversations. Our analysis reveals that, in both pull requests and issues, multi-turn conversations contain three major types of prompts: those that pose follow-up questions (M1), those that introduce the initial task (M2), and those that are refined from a previous prompt (M3). One prompt from DevGPT-PRs and six prompts from DevGPT-Issues were categorized under ``\textit{Others}'' due to their nature of being either casual conversation or lacking sufficient detail to determine their roles.

Below, we describe each category in more detail.

\begin{table}
    \centering
    \caption{Taxonomy of prompts in shared multi-turn conversations}
    \label{tab:Multi-turn_Prompts}
    \begin{tabular}{|l|c|c|}
    \hline
       \textbf{Categories}  & \textbf{PR} & \textbf{Issues} \\
       \hline
      (M1) Iterative follow-up   & 77 (33\%) & 163 (40\%)\\
      (M2) Reveal the initial task   &  62 (26\%) & 118 (29\%)\\
      (M3) Refine prompt   & 40 (17\%) & 56 (14\%)\\
      (M4) Information giving   & 18 (8\%) & 24 (6\%)\\
      (M5) Reveal a new task   & 16 (7\%) & 15 (4\%)\\
      (M6) Negative feedback   & 13 (6\%) & 8 (2\%)\\
      (M7) Asking for clarification   & 9 (4\%) & 19 (5\%)\\ \hline
      Others   & 1  & 6 \\
      % should we put others here? 
      \hline
    \end{tabular}
\end{table}

%This shows that developers often pivot to a new but related task, building upon the context or responses from earlier parts of the conversation. It suggests a dynamic and evolving nature of their inquiry, as they might start with one issue and then move to related aspects based on the information received.

\noindent \textbf{(M1) Iterative follow-up:} In 33\% and 40\% of the prompts in multi-turn DevGPT-PRs and DevGPT-Issues, developers post queries that directly built upon ChatGPT's prior responses or the ongoing context, such as debugging and repairing a solution after code generation by ChatGPT. Such iterative follow-ups typically emerge when the initial task presents a complex challenge that ChatGPT might not fully resolve in a single interaction. Consequently, developers engage in a prompt specifying a follow-up request, enabling ChatGPT to incorporate human feedback and iteratively enhance the proposed solution. 

\noindent \textbf{(M2) Reveal the initial task:} We find that a similar fraction, i.e., 26\% in multi-turn DevGPT-PRs and 29\% in multi-turn DevGPT-Issues, of prompts serve to introduce the initial task to ChatGPT. This distribution highlights that in multi-turn conversations, unlike in single-turn conversations, where the sole prompt is dedicated to outlining the primary task, there are a significant amount of prompts serving other purposes.

\noindent \textbf{(M3) Refine prompt:} Besides iterative follow-up (M1), developers also tend to improve the solution proposed by ChatGPT by providing a refined request prompt with additional context or constraints. The objective is to enhance the response quality for the same query posted in the previous prompt. Refined Prompts account for 17\% of the prompts in multi-turn DevGPT-PRs and 14\% in DevGPT-Issues. 

\noindent \textbf{(M4) Information giving:} In 8\% and 6\% of the prompts in multi-turn DevGPT-PRs and DevGPT-issues, developers do not post any request for ChatGPT, but rather, share knowledge or context with ChatGPT.

\noindent \textbf{(M5) Reveal a new task} We observe that 7\% and 4\% of the prompts in multi-turn DevGPT-PRs and DevGPT-issues are posting a new task to ChatGPT, which is distinct from the task(s) concerned in the prior prompts. This category represents a clear difference from iterative follow-ups (M1), as the new task does not relate to or build upon ChatGPT's prior responses and aims for a different goal. For example, a developer initially requested ChatGPT to generate the SQL corresponding to a Django query set and, in a subsequent prompt, asked for the SQL for a different query set, thereby shifting the focus of the conversation to an entirely new task without prior relevance. 

\noindent \textbf{(M6) Negative feedback:} Within multi-turn conversations, a few (6\% in DevGPT-PRs and 2\% in DevGPT-Issues) prompts contain only negative feedback directed at ChatGPT's previous responses, without providing any information for ChatGPT to improve or further resolve. For instance, ``\textit{Your code is incorrect}'', ``\textit{The same error persists}'', and ``\textit{...does not work}''. This category underscores instances where developers seek to inform ChatGPT of its shortcomings, without seeking further assistance or clarification.

\noindent \textbf{(M7) Asking for clarification:} 4\% and 5\% of prompts in multi-turn DevGPT-PRs and DevGPT-Issues ask ChatGPT to elaborate on its response. These requests for elaboration aim to ensure the comprehensiveness of a solution, e.g., ``\textit{Do I need to do anything else?}''. They also include verification of ChatGPT’s capacity to handle specific tasks, or inquiries to verify whether certain conditions have been considered in the response. Additionally, developers might ask why some alternatives were overlooked by ChatGPT, indicating a deeper engagement with the proposed solutions and a desire to understand the rationale behind ChatGPT's proposed solution.

\subsubsection{RQ 2.2 \rqmultipattern} 
\begin{figure}
    \centering
\includegraphics[width=0.85\linewidth]{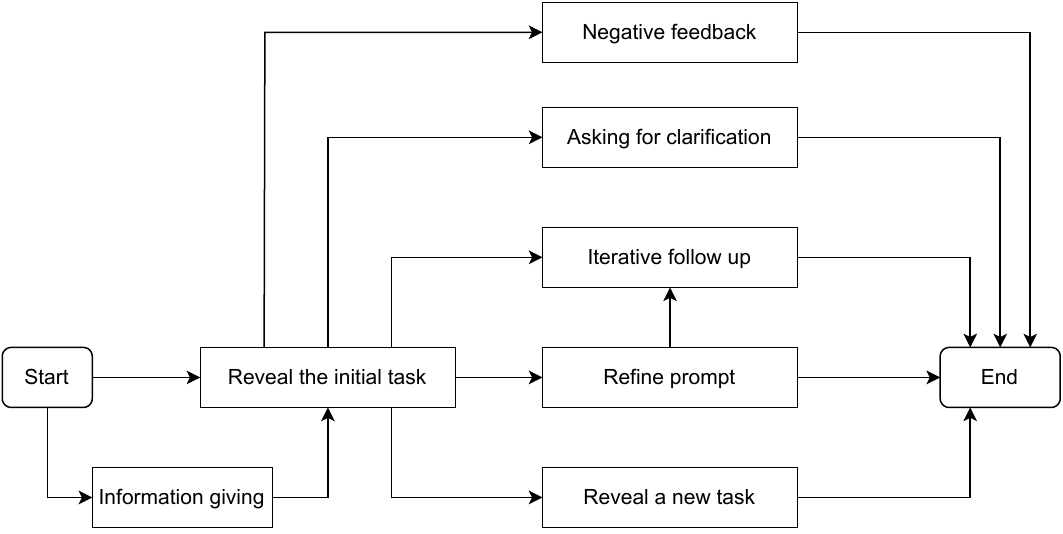}
    \caption{A flow chart presenting common flow patterns in multi-turn conversations.}
    \label{fig:Frequent-Pattern}
\end{figure}

Figure~\ref{fig:Frequent-Pattern} presents the resulting flow chart after applying the postprocessing steps on a Markov Transition Graph based on annotated conversations as a result of RQ2.1. The flow chart applies to multi-turn conversations in both DevGPT-PRs and DevGPT-Issues.

As illustrated in Figure~\ref{fig:Frequent-Pattern}, multi-turn conversations typically start with the presentation of the initial task (M2) or contextual information (M4). Our detailed follow-up analysis reveals that 81\% of multi-turn conversations in DevGPT-PRs and 90\% in DevGPT-Issues begin by outlining the initial task. Conversely, around 13\% of multi-turn conversations in DevGPT-PRs and 3\% in DevGPT-Issues introduce the initial task in the second prompt. In extreme instances, the initial task is disclosed as late as the seventh turn, or, in some cases, the initial task is never explicitly presented—instead, these conversations only present information to ChatGPT without directly stating the task. 

% add the missing pattern 

As for the complete flow, we identified the following patterns based on Figure~\ref{fig:Frequent-Pattern}:

\begin{enumerate}
    \item \( \text{Start}  \rightarrow\text{reveal the initial task} \rightarrow \text{iterative follow up} \rightarrow  \text{end} \)
    \item \(\text{Start }\rightarrow \text{reveal the initial task} \rightarrow \text{refine prompt} \rightarrow  \text{(iterative follow up)}  \rightarrow \text{end}  \)
    \item \( \text{Start} \rightarrow \text{reveal the initial task} \rightarrow \text{reveal a new task } \rightarrow  \text{end} \)
    
    \item \( \text{Start} \rightarrow \text{information giving} \rightarrow \text{reveal the initial task } \rightarrow  \text{...} \rightarrow  \text{end} \)
    %\item \( \text{Start} \rightarrow \text{information giving} \rightarrow \text{iterative follow up } \rightarrow  \text{...} \rightarrow  \text{end} \)
    %\item \( \text{Start} \rightarrow \text{information giving} \rightarrow \text{negative feedback } \rightarrow  \text{...} \rightarrow  \text{end} \)

    \item \( \text{Start} \rightarrow \text{reveal the initial task} \rightarrow \text{asking for clarification} \rightarrow  \text{end} \)
    %\item \( \text{Start} \rightarrow \text{reveal the initial task} \rightarrow \text{asking for clarification} \rightarrow  \text{iterative follow up} \rightarrow  \text{end} \)
    %\item \( \text{Start} \rightarrow \text{reveal the initial task} \rightarrow \text{asking for clarification} \rightarrow  \text{refine prompt} \rightarrow  \text{end} \)
    
    \item \( \text{Start} \rightarrow \text{reveal the initial task} \rightarrow \text{negative feedback} \rightarrow  \text{end} \)
    %\item \( \text{Start} \rightarrow \text{reveal the initial task} \rightarrow \text{negative feedback} \rightarrow  \text{iterative follow up} \rightarrow  \text{end} \)
    %\item \( \text{Start} \rightarrow \text{reveal the initial task} \rightarrow \text{negative feedback} \rightarrow  \text{reveal a new task} \rightarrow  \text{end} \)

\end{enumerate}

Flow patterns (1) to (3) show the most common developer-ChatGPT interaction flows in multi-turn conversations. The initial task is disclosed in the initial prompt, followed by prompts aiming to improve ChatGPT's responses via iterative follow-up, prompt refinements, or to reveal a new task. 

Pattern (4) demonstrates interaction flows started by developers providing information to ChatGPT as the first step. Then, the initial task was revealed, followed by patterns akin to (1) to (3).

Pattern (5) refers to developers asking for clarification from ChatGPT after revealing the initial task and receiving a response from ChatGPT.

Pattern (6) represents an interaction flow in which developers provide negative feedback after revealing the initial task and receiving a response from ChatGPT. 

Although self-loops are excluded from Figure~\ref{fig:Frequent-Pattern}, it's important to note that certain types of prompts, i.e., those revealing a task, refining a previous prompt, posing iterative follow-up questions, or providing information, can occur repetitively. An example of this pattern is observed when a developer successively refines their prompt across two consecutive prompts.

\begin{tcolorbox}[enhanced,width=4.7in,size=fbox,drop shadow southwest,sharp corners]

\textit{RQ2 Summary: We observed similar common patterns in multi-turn conversations shared in GitHub pull requests and issues. Developers engage in iterative conversations with ChatGPT to enhance the quality of the responses. This is achieved by posting follow-up questions or refining a previous prompt to clarify the request or incorporate additional context. Most conversations introduce the initial task within the initial or second prompt, while a few begin by providing ChatGPT with relevant information. We have identified six prevalent flow patterns in multi-turn conversations, which illustrate the dynamic nature of these interactions. } 

\end{tcolorbox}

\section{RQ3: \rqthree}\label{sec:rq3}

%In this research question, we are investigating the characteristics of the sharing behavior of developers of driving developers to share their interactions with ChatGPT in the context of GitHub Issues and Pull Requests (PRs). As per the context of the dataset, it is a common practice among developers to include links to these conversations within their ongoing discussions on GitHub. These links can lead to various sources, ranging from references on StackOverflow to external websites. 

\textbf{Motivation:} In RQ1 and RQ2, we focus on the content in shared conversations. In RQ3, we shift our attention to how and why developers share those conversations in GitHub pull requests and issues. To gain a comprehensive understanding of the characteristics associated with these sharing behaviors within the GitHub ecosystem, we propose three sub-RQs as follows:
\begin{itemize}
    \item \textbf{RQ 3.1 Rationale: \rqpurpose} This question aims to uncover the underlying reasons for developers to share their conversations within GitHub PRs or issues. %Understanding the purposes behind developers' sharing behaviors provides insights into the intentions, goals, and perceived benefits developers associate with sharing ChatGPT conversations in collaborative coding.
    \item \textbf{RQ 3.2 Location: \rqloc} This question examines where the links to shared conversations are posted in GitHub PRs and issues. GitHub PRs and issues contain different components, e.g., description, comment, and code review. %Identifying this characteristic of shared conversations offers insights into the specific sharing locations developers utilize for collaboration and communication.
    \item \textbf{RQ 3.3 Person: \rqwho} This question investigates the involved developers, i.e., the roles and responsibilities of these developers within collaborative development environments like GitHub.
\end{itemize}
 
Answering the above sub-RQs will reveal how developers leverage these shared conversations to enhance their collaboration work within Github. This insight will, in turn, indicate areas where ChatGPT can offer more targeted support to developers in these unique contexts. By understanding these dynamics, we can identify opportunities for ChatGPT and other FM-powered software development supporting tools, to contribute to enhanced productivity and foster stronger collaboration within open-source projects.

\subsection{Approach}
To answer the three sub-RQs, we determine the distribution of PRs and Issues on the combined PR and Issues datasets that are result of the data processing step. Out of 580 total shared conversations, 36.2\% are from the PRs dataset and 63.8\% are from Issues dataset. To achieve a statistically significant sample at a 95\% confidence level and a margin of error of 5\%, we employed stratified random sampling across the PR and Issue categories, selecting a subset of 250 conversations, i.e., 90 from PRs and 160 from Issues. Subsequently, we applied stratified sampling to each of those subsetsj, using the criteria outlined in Table~\ref{tab:Developer_Tasks}. This approach ensured the distribution of software task labels in our samples remained consistent with the findings from RQ1. Then we use the ``MentionedURL'' field in the conversation profile to collect a set of URLs pointing to the PRs and issues containing the selected conversations. This process resulted in the 90 PRs and 160 issues we used for RQ3.

For RQ3.1, similar to RQ2, we use open coding to manually label the rationale behind sharing by checking the usage of the shared conversations within the context of the PRs or issues containing them over three rounds:

\begin{itemize}
    \item In the first round, three co-authors independently labeled 30 cases each from selected PRs and issues (in total 60). Through a follow-up discussion, the annotators developed a coding book containing six distinct labels.   
    \item In the second round, two annotators independently labeled another 20 cases from GitHub PR and GitHub Issues, a total of 40 comments, based on the coding book established in the first round. The two annotators achieved an inter-rater agreement score of 0.78. They then discussed and resolved disagreements and refined the coding book. 
    \item In the third round, the two annotators who participated in the second round continued and equally labeled the remaining cases based on the coding book.
\end{itemize}

For RQ3.2, we manually identified the location of links to shared conversations based on the types of provided content in a PR and issue. Specifically, for PRs, we assess whether the link is present in the PR title, description, comments, or code review comments. For issues, we determine if the link is included in the issue title, description, or comments.

For RQ3.3, we manually identified the role of the developer who shared the conversation. Specifically, we consider PR authors and code reviewers for PRs, and issue assignees, authors, and collaborators for issues.

When we labeled the collected 90 PRs and 160 issues for RQ3.1, we found cases (those labeled as \textit{Others} in Table~\ref{sec:rq3.1}) that could not be utilized in RQ3.2 and RQ3.3. Thus, we ended up with 85 PRs and 154 issues for RQ3.2 and RQ3.3.

\subsection{Results}
In the following paragraphs, we first summarize the distribution of categories for each sub-RQ in Section~\ref{sec:rq3.1} and Section~\ref{sec:rq3.2}, and then present the findings of our cross sub-RQ analysis for RQ3 in Section~\ref{sec:rq3.4}.

\subsubsection{RQ3.1 \rqpurpose} \label{sec:rq3.1}\par

Table~\ref{tab:developer_purpose_share_conversations} details a taxonomy of the derived rationale from analyzing 184 shared conversations and their corresponding PRs and issues. It identifies the top three rationales for sharing conversations in both contexts: (P1) citing the conversation as a potential solution, (P2) referring to the conversation as a source of solution, and (P3) leveraging the conversation to bolster a claim. These rationales collectively represent the majority of sharing instances, accounting for 94\% in selected DevGPT-PRs and 91\% in DevGPT-Issues, respectively. 

Among the 250 analyzed cases, five in DevGPT-PRs and six in DevGPT-Issues were categorized as \textit{Others} due to challenges in identifying their specific purpose. These challenges arose from several issues: 1) the removal of the link to the shared conversation, 2) the deletion of the linked pull request or issue, or 3) the content within the pull request or issue is in a language other than English. These factors contributed to the difficulty in discerning the intended use or contribution of the shared conversation in these instances.

In one PR and 15 issues, developers only provided the URLs to the shared conversations without any explanation, forming the ``\textit{Direct link}'' category.

Below, we describe each category in more detail. We use ``LINKTOCHAT'' to refer to the omitted links to the actual shared conversation. 

\begin{table}[h]
    \centering
    \caption{Taxonomy of developer's purpose of sharing.} \label{tab:developer_purpose_share_conversations}
    \begin{tabular}{|p{7cm}|c|c|c|}
    \hline
       \textbf{Category}  & \textbf{PR} & \textbf{Issue}  \\
    \hline
      With clear purpose: & 84 (100\%)  & 139 (100\%)\\
      (P1) Reference to a source of solution   & 31 (37\%) &  30 (22\%)  \\
      (P2) Reference to a potential solution   & 28 (33\%) & 74 (53\%) \\
      (P3) Support a claim  & 20 (24\%) & 23 (17\%)  \\
      (P4) Answer a question  & 3 (4\%) & 4 (3\%)  \\
      (P5) Illustrate an example   & 2 (2\%) & 8 (6\%)\\
      \hline
      Direct link & 1  & 15  \\
      \hline
      Others & 5  & 6  \\
    \hline
    \end{tabular}
\end{table}

\noindent \textbf{(P1) Reference to a source of solution:} Developers reference shared conversations as the source of solutions in 37\% of the cases within selected pull requests (with clear purpose), significantly more prevalent than in issues, where it accounts for 22\%. This purpose often occurs when developers initiate a GitHub PR or issue following guidance or solutions suggested by ChatGPT. For instance, ``\textit{Current solution is based on LINKTOCHAT. Now it seems work}''. This observation indicates that ChatGPT has supported developers with their contributions to open-source projects. Furthermore, it reveals that developers are inclined to include additional context and details about their contributions by referencing their interactions with ChatGPT, which will enrich the collaborative development process by providing transparent insight into the reasoning and evolution of their contributions.

\noindent \textbf{(P2) Reference to a potential solution:} Beyond referencing solutions for already implemented code (P1), developers also refer to shared conversations as potential solutions. In pull requests, 33\% of cases fall into this category, but it is significantly more prevalent in issues, accounting for 53\%. The higher occurrence in issues is potentially due to their nature as places for exploring and debating prospective solutions aimed at issue resolution. In pull requests, we still find P2 a common category because shared conversations are leveraged to propose potential solutions for addressing code reviews, as exemplified in Figure~\ref{fig:example}. 

\noindent \textbf{(P3) Support a claim:} Developers shared conversations to support their claims, arguments, or suggestions in 24\% of PRs and 17\% of identified issues with a clear purpose. For instance, ``\textit{I have doubts about the utility of soft delete ... i consulted the chatgpt oracle, and it seems to not like soft deletion too: LINKTOCHAT}''. Such instances demonstrate the strategic use of ChatGPT as an authoritative source or ``oracle'' to lend weight to a developer's perspective or to resolve debates within the development process.

\noindent \textbf{(P4) Answer a question:} In 4\% and 3\% of the cases in PRs and issues, developers use the shared conversation as an answer to a question raised in the previous comment. For instance, a code reviewer in a pull request asked ``\textit{This may need to be changed in transifex?...}''. The pull request author answered: ``\textit{I think it does sync. LINKTOCHAT}''.

\noindent \textbf{(P5) Illustrate an example:} In a few cases, one for PR and eight for the issue, developers use the shared conversation as an example to demonstrate a concrete example of their mentioned concepts. For instance, ``\textit{We want a full exhaustive hierarchical enumeration of all the tasks a person might do on a computer. Similar to LINKTOCHAT but more exhaustive}''.

%\noindent \textbf{Unclear}, developers only share the link to the shared chat and no additional text or explanation was provided in the PR or Issue comment.

%\textbf{Non-english}, Pull Request or Issue conversations that are not in English. 

%\textbf{Removed}, the linked Pull Request or Issue was removed. 

%\textbf{Unsure}, Pull Request or Issues we are not sure which category it belongs to. 

\subsubsection{RQ3.2 \rqloc}\label{sec:rq3.2}\par

Table~\ref{tab:location} shows the distribution of locations of links to shared conversations in 85 pull requests and 154 issues. %The equitable distribution of shared conversation links within pull requests indicates no particular preference for their placement. In contrast, issues demonstrate a pronounced tendency for shared conversations to be located in comments rather than in descriptions or titles.

\begin{table}[h]
    \centering
    \caption{Locations of shared conversations.} \label{tab:location}
    \begin{tabular}{|l|c|c|}
    \hline
       \textbf{Location}  & \textbf{Count (\%)}  \\
    \hline
      \textbf{In pull requests:} &  85 (100\%) \\
      \hspace*{1em} Code review comment  & 30 (35\%)  \\
      \hspace*{1em} Pull request comment   & 29 (34\%)  \\
      \hspace*{1em} Pull request description  & 26  (31\%)  \\
      \hline
      \textbf{In issues:} & 154 (100\%)  \\
      \hspace*{1em} Issue comment  & 97 (63\%)   \\
      \hspace*{1em} Issue description   & 56 (36\%)  \\
      \hspace*{1em} Issue title  & 1 (1\%)  \\
    \hline
    \end{tabular}
\end{table}

\subsubsection{RQ3.3 \rqwho}\label{sec:rq3.3}\par Table~\ref{tab:who} shows the distribution of locations of links to shared conversations in 85 pull requests and 154 issues. %The numbers reveal that: in pull requests, the likelihood of PR authors and reviewers sharing conversations is equally distributed, indicating a collaborative engagement in the sharing process. Conversely, within issues, it is predominantly the issue authors who post links to shared conversations. 

\begin{table}[h]
    \centering
    \caption{The roles of developers who shared conversations.} \label{tab:who}
    \begin{tabular}{|l|c|c|}
    \hline
       \textbf{Location}  & \textbf{Count (\%)}  \\
    \hline
      \textbf{In pull requests:} &  85 (100\%) \\
      \hspace*{1em} PR author  & 45 (53\%)  \\
      \hspace*{1em} Code reviewer   & 40 (47\%)  \\
      \hline
      \textbf{In issues:} & 154 (100\%)  \\
      \hspace*{1em} Issue author  & 109 (71\%)   \\
      \hspace*{1em} Issue collaborator   & 39 (25\%)  \\
      \hspace*{1em} Issue assignee  & 6 (4\%)  \\
    \hline
    \end{tabular}
\end{table}

%The visualization of the relationships in GitHub Pull Request comments is demonstrated by Figure \ref{fig:sankey-pr}. The diagram for relationships in GitHub Issues is demonstrated by Figure \ref{fig:sankey-issue}. In these diagrams, parallel categories are employed. Two sets of parallel categories were defined: "Person - Purpose" and "Person - Location". 

\subsubsection{Cross Sub-RQ Findings.} \label{sec:rq3.4}
To further understand the sharing behaviors of developers, we conducted an in-depth analysis to investigate the relationships between three considered aspects, i.e., rationale, location, and person. 

\begin{figure}[h]
    \centering
    \vspace{-0.5cm}
    \includegraphics[width=0.86\linewidth]{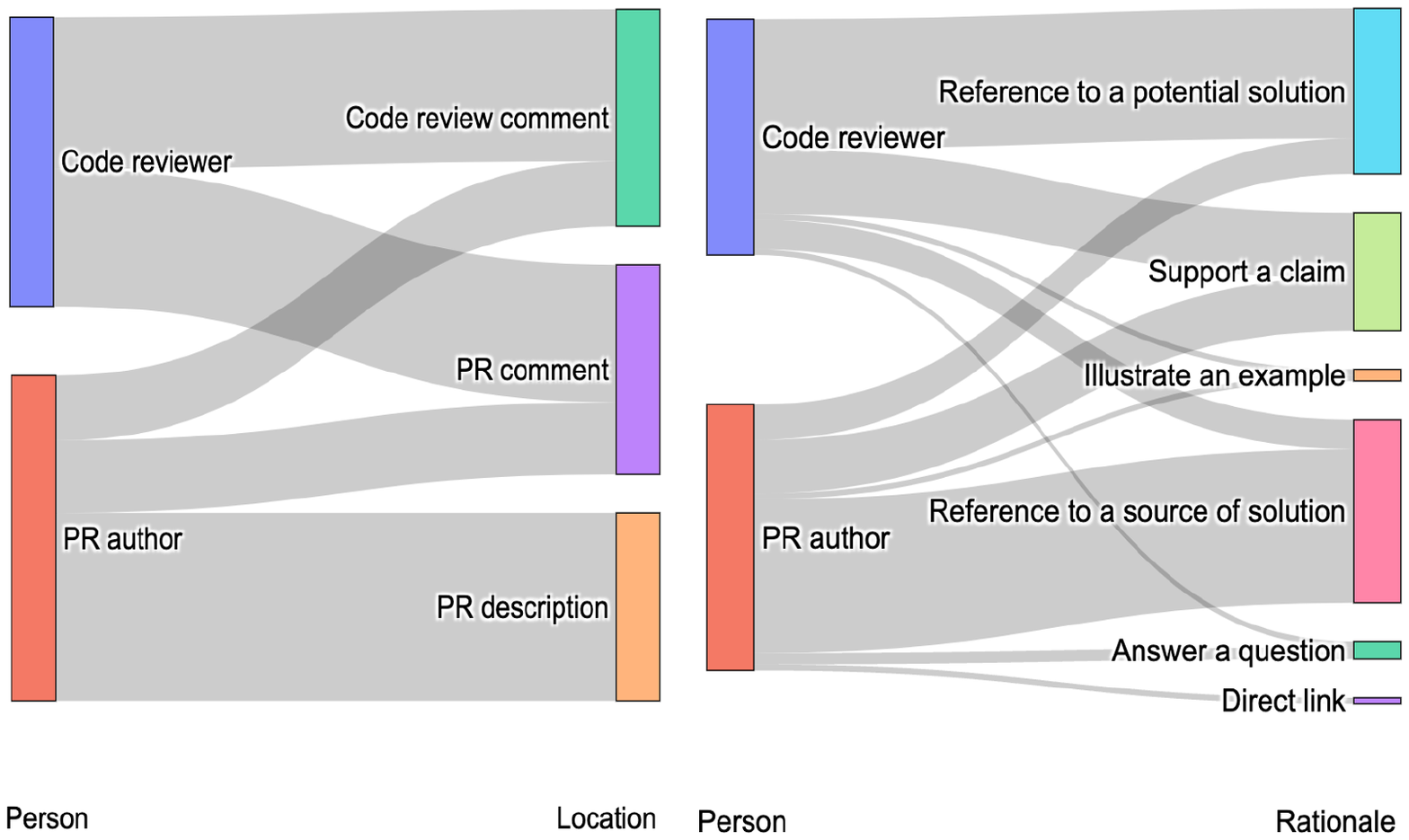}
    \caption{The diagram presented on the left illustrates the specific location at which the individual shares the ChatGPT conversation in Github PRs. Conversely, the diagram displayed on the right delineates the rationale underlying the individual's decision to share the ChatGPT conversation in Github PR.}
    \label{fig:sankey-pr}
\end{figure}

Figure~\ref{fig:sankey-pr} shows two Sankey plots (``\textit{Person-Location}'' and ``\textit{Person-Rationale}'' indicating the behavior patterns (where they share and why they share) of developers based on their roles in the pull requests. The width of the lines connecting these sets corresponds to the frequency of co-occurrence between the connected variables. In the \textit{Person-Location} plot, we observe that reviewers share conversations in code review comments and pull request comments with almost equal probability, 53\%, and 47\%, respectively. Authors usually share their conversations in the pull request description (58\%) and in the pull request comment and code review comment with similar probability, 22\% and 20\%, respectively. In the \textit{Person-Rationale} plot, we observed that reviewers usually share their conversations as a reference to a potential solution (P2) (55\%), then to support a claim (P3) (28\%), and last as a reference to a source of solution (P1) (13\%). Authors show a different sharing behavior; they mostly use shared conversations as a source of solution (P1) (58\%), then support a claim (P3) (20\%), and as a potential solution (P2) (13\%).

%\huizi{is below paragraph needed?}
%To improve the clarity and detail of your description of the analysis presented in Figure~\ref{fig:sankey-issue}, focusing on the distribution and purposes of shared conversations in issues, consider the following enhanced version:

\begin{figure}[h]
    \centering
    \includegraphics[width=0.86\linewidth]{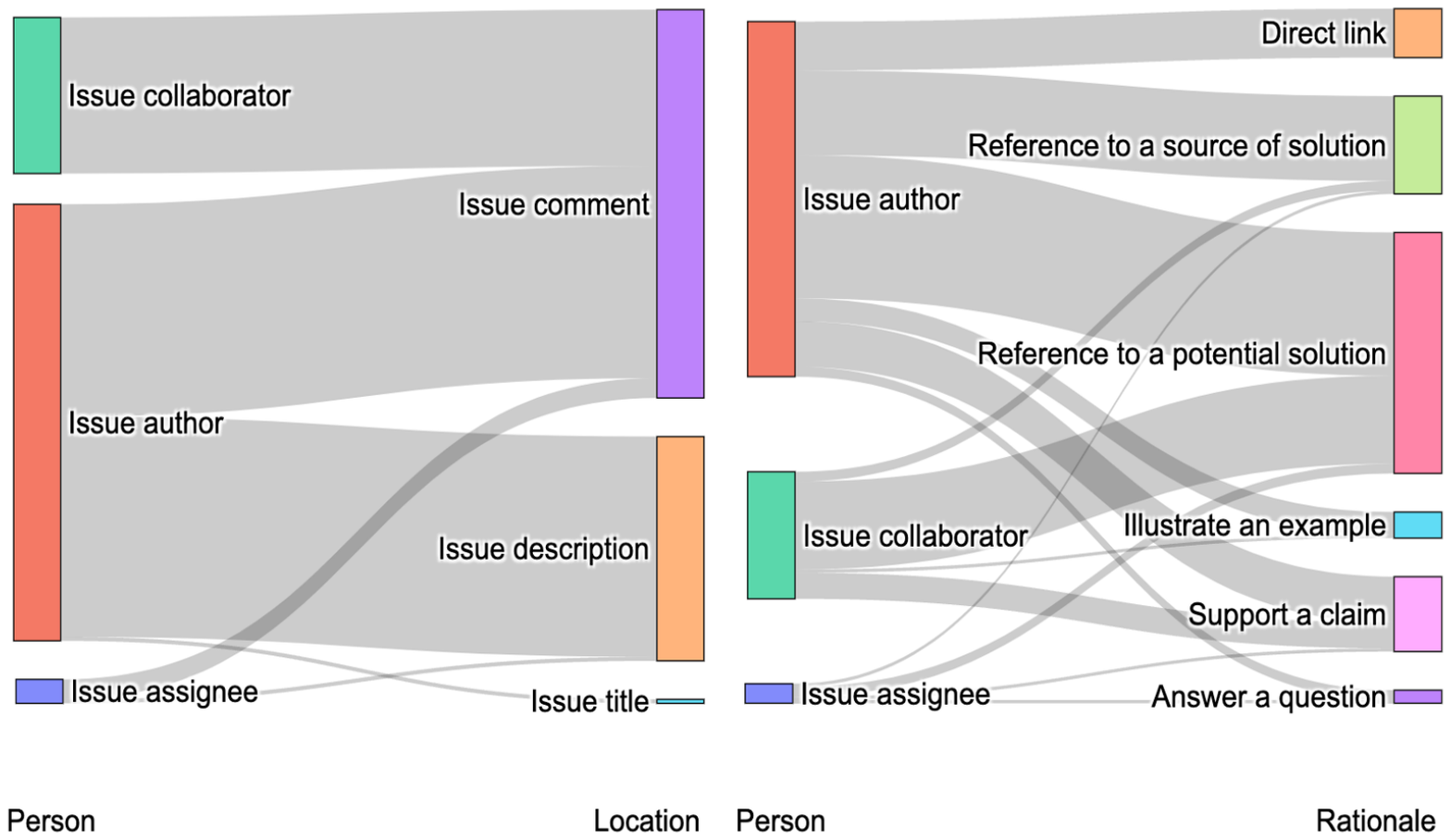}
    \vspace{-0.5cm}
    \caption{The diagram presented on the left illustrates the specific location at which the individual shares the ChatGPT conversation in Github Issues. Conversely, the diagram displayed on the right delineates the rationale underlying the individual's decision to share the ChatGPT conversation in Github Issues.}
    \label{fig:sankey-issue}
\end{figure}

Figure~\ref{fig:sankey-issue} provides a detailed analysis of shared conversations within issues, employing a \textit{Person-Location} and a \textit{Person-Rationale} perspective. The \textit{Person-Location} analysis reveals that collaborators and assignees exclusively share conversations in issue comments. Issue authors (reporters) predominantly share conversations in issue comments and descriptions, with a nearly even split of 49\% for comments and 51\% for descriptions. One issue author even includes the link directly in the issue title. The \textit{Person-Rationale} plot further dissects the motivations behind sharing conversations for different types of developers in issues. Issue authors primarily use shared conversations to highlight them as potential solutions (P2), counting for 41\% of their shares. This is followed by references to conversations as sources of solutions (P1) and support for claims (P3), which are shared with similar probabilities of 24\% and 13\%, respectively. Collaborators, on the other hand, are more inclined to share conversations as a potential solution (P2) in 69\% of cases and, to a lesser extent, to support a claim (P3) at 21\%. Assignees prefer sharing conversations as potential solutions (P2) in 50\% of instances, with the remaining equally referencing other rationales (P1, P3 and P4) at 17\% respectively.

The observed patterns of developers' sharing behaviors reflect the distinct dynamics within pull requests and issues. In pull requests, authors are responsible for crafting comprehensive descriptions, whereas reviewers engage primarily through code review and pull request comments. This distinction is mirrored in their motivations for sharing links to conversations: authors often share to cite the origin of their implemented solution, while reviewers tend to share to introduce or suggest alternative solutions to the author. In issues, the behavior diverges; authors actively contribute to the issue's initial description and subsequent comments. Collaborators and assignees, in contrast, limit their contributions to comments. This dynamic sees authors using shared conversations as evidence supporting their proposed solutions and substantiating their claims. Collaborators, on the other hand, frequently share conversations to present potential solutions or to bolster a claim.

Our observations underscore the role of shared conversations in facilitating collaborative problem-solving, with each participant leveraging these exchanges according to their role and the unique collaborative needs of pull requests and issues.

\begin{tcolorbox}[enhanced,width=4.7in,size=fbox,drop shadow southwest,sharp corners]

\textit{RQ3 Summary: Developers mainly share their conversations with ChatGPT within GitHub PRs and issues to cite these conversations as either the foundation of a solution, a prospective solution, or to support their claims. Developers leverage shared conversations to augment their role-specific inputs, effectively employing these exchanges to enhance clarity, propose alternatives, or reinforce arguments for more efficient and transparent collaborative software development.} 

\end{tcolorbox}

\section{Discussions} \label{sec:disc}

\noindent \textbf{Implications for Designing and Investigating FM-powered SE collaboration tools.} The most important finding from our study is that developers do share their conversations with ChatGPT while contributing to open-source projects. This insight opens a new view for researchers and FM practitioners assessing the role and influence of FM-powered software development tools, such as ChatGPT, within the realm of collaborative coding. It underscores the potential of these tools to not only assist individual developers but also to enhance the collective productivity and innovation of open-source communities.

Furthermore, our study provides several taxonomies that researchers can further utilize to characterize developers' interactions with ChatGPT or other FM-powered software development tools. For instance, the taxonomy and annotated prompts in RQ1 can be leveraged to develop a learning-based approach that can automatically identify tasks per interest and analyze the corresponding response quality. Designers can also leverage our reported frequency of software engineering tasks to prioritize improvement for their tools. The answers to RQ3 reveal how developers with different roles use shared conversations with ChatGPT in collaborative coding, which can be used to design FM-powered tools tailored to support developers with other roles.

\vspace{0.1cm}
\noindent \textbf{Implications for Benchmarking FM for SE tasks.}
Our findings from RQ1 shed light on future benchmark designs for evaluating the impact of FMs in different types of software engineering tasks. In RQ1, we find multiple types of input for code generation and issues resolving inquiries, but those types are not fully captured by existing benchmarks. For instance, the widely recognized code generation benchmark, HumanEval~\citep{chen2021evaluating}, relies on textual specifications and method signatures. Yet, our analysis shows that nearly half of the code generation prompts (47\%) include initial code drafts alongside textual descriptions. Similarly, our examination of prompts categorized under \textit{(C4) Issue resolving} indicates that a significant portion (36\%) of issue resolution requests involve sharing error messages or execution traces, often without accompanying source code. Therefore, we recommend that researchers designing future benchmarks take these findings into account.

Our observation that multi-turn conversations are often utilized also motivates future evaluation of FMs allowing multi-turn interactions. Currently, there are only a few studies allowing multi-turn code generation~\citep{wang2024executable, nijkamp2022codegen}. 

%but other forms of refinement and iterative approaches are still unexplored. 

Last but not least, we observed many other tasks beyond code generation and issue resolution, such as code review, conceptual question, and documentation, which are rarely considered as benchmark tasks for FM-powered software development tools. 
%\marcos{Are we saying that they should be considered as an implication to SE benchmarking?}

\vspace{0.1cm}
\noindent \textbf{Implications for Prompt Engineering.} The findings from RQ2 highlight the frequent use of multi-turn strategies to improve ChatGPT's solutions iteratively. The flow chart shown in Figure~\ref{fig:Frequent-Pattern} illustrates the diverse approaches developers employ in these interactions. This finding motivates future investigations into the efficiency of developers' prompting techniques within these multi-turn conversations. Specifically, whether the best practices in prompt engineering have been applied and whether improved prompts can effectively alter the flow of these interactions is a future direction for enhancing the utility and effectiveness of FM-powered tools in software development. 

%Our observations from examining comments where the developer thinks ChatGPT hallucinated in the response while the developer only prompted ChatGPT once, and the prompt did not follow the best practices of prompt engineering. During our investigation of developers' shared multi-turn conversations, iterative follow up and refine prompts occurs frequently. By in-depth analysis of examples from these prompts, we identify that developers do not always follow the best practice of prompting in multi-turn conversations either. A prompting guide for each category of software engineering tasks can be developed to better help developers in seeking help from ChatGPT.   

\section{Threats to Validity} \label{sec:threats}
\noindent\textbf{Internal Validity:} 
Open coding was employed in our study as the major methodology we conducted. This approach introduces subjectiveness to our results as the annotators may hold different understandings of the coding schema. To mitigate this threat, we follow below best practices below: (1) we conduct multiple rounds of individual coding, and we ensure the Cohen’s Kappa score reflecting the agreement between the annotators is at least substantial; (2) all disagreement is discussed with all annotators until agreement reached; (3) two rounds of revisit on the entire label-set are performed after individual labeling to mitigate any miscoded. 

Moreover, there were cases where their labels were difficult to decide as they lay within the boundaries of two categories. As we employed a single label for all data, these boundary cases were labeled with a single label that the annotators thought was the best label to describe them.

\noindent\textbf{External Validity:}
Our results are derived from the specific dataset, DevGPT, where only developers shared conversations in GitHub Pull Request and Issues are considered. Furthermore, the dataset only contains data collected up to October 2023. Thus, our results may not generalize to all Developer-ChatGPT conversations. Developers shared conversations and usage of these shared conversations existing in other open-source platforms might vary. Developers shared conversations and usage in general contexts, i.e., when these shared conversations are not posted in GitHub PRs and issues may not follow the findings discussed in this paper. We encourage researchers to reproduce our study on a larger dataset or data collected from other open-source communities.

\noindent\textbf{Construct Validity:}
To ensure our taxonomy presented in this study best captures the data, we develop our coding book based on previous studies and expand on them to include ChatGPT-specific scenarios. We build our coding book for RQ1 based on the taxonomy proposed by Beyer et al.~\citeyear{beyer2020kind} and Hata et al. ~\citeyear{hata2022github}. Our RQ2 coding book is expanded on the taxonomy proposed by Huang et al.~\citeyear{huang2018automating} and Qu et al.~\citeyear{qu2019user}.

%When conducting  Open Coding labeling, we noticed that some developer-ChatGPT conversations posted in GitHub Pull Request and Github Issues were deleted by developers and some GitHub Pull Request and Github Issues containing the shared conversations were removed. Several reasons may contribute to the removal of these conversations, pull requests, and issues. One possible reason is that the developer is not satisfied with the developer-ChatGPT conversation and removed it from the history chat list. In these cases, we might employ different labels for these cases.

\vspace{-0.4cm}
\section{Related Work}\label{sec:related}

In this section, we first present related work on the usability of FM-powered tools (Section~\ref{sec:related_fm}). As our study analyzes shared conversations in GitHub issues and pull requests, we also present existing studies on analyzing conversations related to software engineering (Section~\ref{sec:related_conversation}) and link sharing in SE platforms, i.e., GitHub and Stack Overflow (Section~\ref{sec:related_sharing}).

\subsection{Understanding Developers' Interactions with FM-powered Software Development Tools} \label{sec:related_fm}

Several studies investigated the interactions between FM-powered software development tools and developers. These studies adopted either user studies or surveys as their methodology, with a significant emphasis on GitHub Copilot, a code completion tool powered by OpenAI's Codex model. 

User studies, such as those conducted by Vaithilingam et al.~\citeyear{vaithilingam2022expectation}, Mozannar et al.~\citeyear{mozannar2022reading}, Barke et al.~\citeyear{barke2023grounded} and Ross et al.~\citeyear{ross2023programmer}, were performed to understand how developers interact with GitHub Copilot. Vaithilingam et al.~\citeyear{vaithilingam2022expectation} reported a general preference among developers for incorporating Copilot into their daily programming tasks via a user study with 24 participants. However, they also found that developers face challenges in understanding and debugging the code that Copilot generates. Mozannar et al.~\citeyear{mozannar2022reading} developed a taxonomy of 12 programmer activities associated with GitHub Copilot use, such as verifying suggestions and debugging. Their analysis of 21 developers' interactions with Copilot revealed that a substantial portion of coding sessions were spent on activities like double-checking and editing Copilot's suggestions. Barke et al.'s user study~\citeyear{barke2023grounded} with 20 participants reported a bimodal usage pattern among developers: they employ Copilot both to speed up familiar tasks (``acceleration mode'') and to explore solutions for new challenges (``exploration mode''). Moreover, programmers may defer thoughts on suggestions for later and simply accept them when displayed. Ross et al.~\citeyear{ross2023programmer} conducted a user study with 42 developers with follow-up surveys using their proposed program assistant, which relies on OpenAI's codex API to support code edit suggest and chat function. They found that 83\% of participants rated the importance of the ability to ask follow-up questions as being ``somewhat'' or ``a great deal''.

On the survey front, large-scale studies by Ziegler et al.~\citeyear{ziegler2022productivity} and Liang et al.~\citeyear{liang2024large} explored developer perceptions of productivity and usability challenges with AI programming assistants, including GitHub Copilot. Ziegler et al. (researchers from GitHub) collected survey responses from 2,631 developers with IDE usage data. They identified the identifying the acceptance rate of Copilot's suggestions as a significant predictor of perceived productivity. Interestingly, acceptance rates varied across programming languages, i.e., 23.3\%, 27.9\%, and 28.8\% of Copilot’s suggestions were accepted for TypeScript, JavaScript, and Python, respectively, and 22.2\% for all other languages. More recently, Liang et al.~\citeyear{liang2024large} performed another survey with 410 developers with a focus on the usability challenges of many AI programming assistants, including GitHub Copilot and ChatGPT. Their results show developers are most motivated to use AI programming assistants because they help developers reduce key-strokes, finish programming tasks quickly, and recall syntax. They also found that the most frequent usability challenges were understanding how input leads to output code and how to control the tool's generations.

Similar to prior studies, our study aims to explore the potentials and limitations of FM-powered tools, i.e., ChatGPT, in software development by analyzing the interaction between developers and ChatGPT. But different from them, we explore ChatGPT's role in supporting developers' collaboration in coding on GitHub by analyzing developers' shared conversations with ChatGPT in GitHub issues and pull requests. We deployed a mixed research method, including qualitative and quantitative studies, rather than using user studies or user surveys.

\subsection{Conversation Analysis in Software Engineering} \label{sec:related_conversation}

In software development, conversations play a crucial role in facilitating collaboration, knowledge sharing, and problem-solving among developers. These interactions vary widely, ranging from email lists to issue discussions on open-source software (OSS) platforms like GitHub to technical queries on community forums such as Stack Overflow. Each type of conversation has unique characteristics and serves distinct purposes within the software development lifecycle. 

Di Sorbo et al.~\citeyear{di2015development} manually annotated 100 emails taken from the Qt project development mailing list and proposed a taxonomy that categorized communications between developers into six types, including feature request, opinion asking, problem discovery, solution proposal, information seeking, and information giving. Building upon this framework, Huang et al.~\citeyear{huang2018automating} sampled 5,408 sentences from comments recorded in issue tracking systems of four large and popular projects hosted on GitHub. Their manual categorization refined Di Sorbo et al.’s taxonomy by adding new categories, such as ``aspect evaluation'' and ``meaningless (less informative)''. Besides general categorization, there are also studies that focus on specific topics from conversations. Arya et al.~\citeyear{arya2019analysis} also investigated issue discussions, but with a focus on fine-grained information types mentioned in conversations. They identified 16 distinct types of information and compiled a labeled corpus of 4,656 sentences. Nurwidyantoro et al.~\citeyear{nurwidyantoro2022human} annotated 1,097 issue discussions collected from three Android projects to understand human values in software development artifacts. Their results show that value themes could be found in issue discussions (33\% of the inspected issues). Beyond issues, studies have also explored conversations in pull requests and online chat rooms. Viviani et al.~\citeyear{viviani2019locating} proposed an automated solution to locate the points of the discussion where developers discuss design in pull requests. Shi et al. ~\citeyear{shi2021first} conducted a comprehensive empirical study on developers’ live chat in gitter, exploring interaction dynamics, community structures, discussion topics, and interaction patterns. In contrast to existing research that primarily focuses on developer-to-developer communication, our study focuses on developer-to-tool interactions. We introduce two new taxonomies: one that categorizes SE-related inquiry types and another that characterizes the role of sentences within multi-turn conversations, drawing inspiration from literature for certain categories like ``information giving''.

Our findings in RQ1 highlight that most shared conversations with ChatGPT in GitHub issues and pull requests revolve around seeking assistance with SE tasks, i.e., these conversations are predominantly information-seeking conversations. Before the advent of FM-powered tools in software development, such inquiries were commonly posed on Q\&A sites like Stack Overflow (SO). In the literature, analyzing questions developers post on SO is a common practice to understand developers' faced challenges in a specific domain, such as mobile development~\citep{rosen2016mobile}, blockchain~\citep{wan2019discussed}, and deploying deep learning application~\citep{zhang2019empirical}. Yet, the studies most relevant to our research are those aimed at categorizing the general types of questions posed by developers on SO. Treude et al.~\citeyear{treude2011programmers} were the first ones investigating the question categories of posts of SO. In 385 manually analyzed posts, they found 10 question categories: how-to, discrepancy, environment, error, decision help, conceptual, review, non-functional, novice, and noise. Similarly, Rosen and Shihab~\citeyear{rosen2016mobile} manually categorized 384 posts of SO for the mobile operating systems Android, Apple, and Windows, each into three main question categories: How, What, and Why. More recently, Beyer et al.~\citeyear{beyer2020kind} have advanced this line of work by manually classifying 1,000 SO questions, proposing a new taxonomy that amalgamates all previously identified question categories. This taxonomy comprises seven high-level categories: API usage, conceptual issues, discrepancies, errors, reviews, API changes, and learning. Our taxonomy in RQ1, inspired by Beyer et al., aligns with many SE-related inquiries observed on SO but also distinguishes the unique types of inquiries directed at ChatGPT, such as those related to documentation improvement, code comprehension, data generation, and data formatting. Moreover, we discuss the role of sentences (developers' prompts) in multi-turn conversation and developers' sharing behavior, which is not covered in prior studies.

\subsection{Linking Sharing in GitHub and Stack Overflow} \label{sec:related_sharing}

Link sharing serves as a pivotal method for knowledge sharing within developer communities, particularly on Q\&A sites like Stack Overflow (SO) and social coding platforms such as GitHub. This practice, widely recognized for facilitating the exchange of insights about software development tools and libraries, underpins much of the collaborative ethos in these communities.

There has been extensive research on developers' link-sharing behaviors on Stack Overflow. Gómez et al.~\citeyear{gomez2013study} observed that a considerable portion of the shared links on SO are aimed at spreading knowledge about new software development tools and libraries. Ye et al.~\citeyear{ye2017structure} further explored this phenomenon, examining the structure and dynamics of SO's knowledge network through link sharing. Their findings highlight that developers share links for a variety of reasons, with referencing information for problem-solving being the most common purpose. Baltes et al.~\citeyear{baltes2020contextual} provided an in-depth analysis of how and why documentation links are cited within SO posts by examining 759 shared links to understand their roles and the importance of context in interpreting these references. More recent studies by Liu et al.~\citeyear{liu2021broken,liu2022exploratory} have analyzed the characteristics of broken shared links and the patterns of repeatedly referenced links within SO posts.

Beyond Stack Overflow, research has extended to the use of links within GitHub's ecosystem, particularly in issues and pull requests. Zampetti et al.~\citeyear{zampetti2017developers} explored the extent and purposes behind developers' references to external online resources in pull requests, indicating a strong inclination towards acquiring new knowledge or addressing specific problems. Zhang et al.~\citeyear{zhang2018within} observed that developers tend to link more cross-project or cross-ecosystem issues over time. Li et al.~\citeyear{li2018issue} conducted a comprehensive study to understand why developers create links within issues and pull requests on GitHub and how these links impact software development. They manually identified six types of relationships in linking behavior: dependent, duplicate, relevant, referenced, fixed, and enhanced. Hata et al.~\citeyear{hata20199} collected 9.6 million links in source code comments. They found that over 80\% of repositories contain at least one link, pointing to the prevalence of link sharing for purposes such as providing metadata or attributions. Wang et al.~\citeyear{wang2021understanding} emphasized the critical role of shared links in code review, demonstrating how they serve as vital resources for both authors and review teams. Xiao et al.~\citeyear{xiao202318} highlighted the issue of link decay in commit messages and the various purposes for which developers include links, primarily for context provision.

Our results in RQ3 align with findings from prior studies. For instance, similar to these investigations, we identified instances where shared links, specifically those encapsulating dialogues between developers and ChatGPT, were broken. In response to this issue, we observed a practice among some developers who, instead of solely sharing a link, opted to embed the entire conversation within issue comments or pull requests as a form of reference (when manually analyzing PRs and issues in RQ3). We also found the shared links play an important role in supporting collaboration in coding. Unlike the predominant trend observed in SO, where documentation links are frequently cited, or in GitHub, where cross-project and ecosystem software artifact links prevail, our study sheds light on the unique nature of links to ChatGPT conversations. These links represent a novel vector for knowledge sharing and collaboration.

\section{Conclusion and Future Work}\label{sec:conclusion}

In this paper, we study the role of ChatGPT in collaborative coding by analyzing developers' shared conversations with ChatGPT in GitHub pull requests and issues, leveraging the DevGPT dataset.

Our key findings include: (1) Developers seek ChatGPT's assistance across 16 types of software engineering inquiries. The most frequently encountered requests involve code generation, conceptual understanding, how-to guidance, issue resolution, and code review. (2) In code generation and issue resolution tasks, developers often go beyond the conventional inputs of textual descriptions or buggy code, which are standard benchmarks for FMs in code generation and program repair. This indicates a broader range of inputs being utilized in real-world scenarios. (3) Developers engage with ChatGPT during multi-turn conversations through iterative follow-up questions, prompt refinement, and clarification inquiries. These methods are employed to enhance the quality and relevance of ChatGPT's responses progressively. (4) Developers with different roles—such as issue authors, PR authors, and code reviewers—utilize shared conversations with ChatGPT to supplement their role-specific contributions. This practice aims to improve the efficiency and transparency of collaborative software development processes.

In the future, we plan to propose automated approaches that can automatically identify the types of prompts based on our taxonomies. We also plan to explore whether the existing best practices for prompt engineering have been applied in the collected shared conversations and if applying them will influence the flow of multi-turn conversations.

\begin{acknowledgements}
We acknowledge the support of the Natural Sciences and Engineering Research Council of Canada (NSERC), [funding reference number: RGPIN-2019-05071].
\end{acknowledgements}

\section*{Conflict of Interest}
The authors declare that they have no conflict of interest.

\section*{Data Availability Statements}
The results, source code, and data related to this study are available at \url{https://github.com/RISElabQueens/analyzing-shared-conversation}

%\pagebreak 

%\bibliographystyle{apalike}      % sadegh: as the basic style doesn't work properly
\bibliographystyle{spbasic}      % basic style, author-year citations

\bibliography{references.bib}

\begin{thebibliography}{45}
\providecommand{\natexlab}[1]{#1}
\providecommand{\url}[1]{{#1}}
\providecommand{\urlprefix}{URL }
\expandafter\ifx\csname urlstyle\endcsname\relax
  \providecommand{\doi}[1]{DOI~\discretionary{}{}{}#1}\else
  \providecommand{\doi}{DOI~\discretionary{}{}{}\begingroup \urlstyle{rm}\Url}\fi
\providecommand{\eprint}[2][]{\url{#2}}

\bibitem[{Arya et~al.(2019)Arya, Wang, Guo, and Cheng}]{arya2019analysis}
Arya D, Wang W, Guo JL, Cheng J (2019) Analysis and detection of information types of open source software issue discussions. In: 2019 IEEE/ACM 41st International Conference on Software Engineering (ICSE), IEEE, pp 454--464

\bibitem[{Baltes et~al.(2020)Baltes, Treude, and Robillard}]{baltes2020contextual}
Baltes S, Treude C, Robillard MP (2020) Contextual documentation referencing on stack overflow. IEEE Transactions on Software Engineering 48(1):135--149

\bibitem[{Barke et~al.(2023)Barke, James, and Polikarpova}]{barke2023grounded}
Barke S, James MB, Polikarpova N (2023) Grounded copilot: How programmers interact with code-generating models. Proceedings of the ACM on Programming Languages 7(OOPSLA1):85--111

\bibitem[{Beyer et~al.(2020)Beyer, Macho, Di~Penta, and Pinzger}]{beyer2020kind}
Beyer S, Macho C, Di~Penta M, Pinzger M (2020) What kind of questions do developers ask on stack overflow? a comparison of automated approaches to classify posts into question categories. Empirical Software Engineering 25:2258--2301

\bibitem[{Chen et~al.(2021)Chen, Tworek, Jun, Yuan, Pinto, Kaplan, Edwards, Burda, Joseph, Brockman et~al.}]{chen2021evaluating}
Chen M, Tworek J, Jun H, Yuan Q, Pinto HPdO, Kaplan J, Edwards H, Burda Y, Joseph N, Brockman G, et~al. (2021) Evaluating large language models trained on code. arXiv preprint arXiv:210703374

\bibitem[{Deng et~al.(2024)Deng, Xia, Yang, Zhang, Yang, and Zhang}]{deng2024large}
Deng Y, Xia CS, Yang C, Zhang SD, Yang S, Zhang L (2024) Large language models are edge-case generators: Crafting unusual programs for fuzzing deep learning libraries. In: Proceedings of the 46th IEEE/ACM International Conference on Software Engineering, pp 1--13

\bibitem[{Di~Sorbo et~al.(2015)Di~Sorbo, Panichella, Visaggio, Di~Penta, Canfora, and Gall}]{di2015development}
Di~Sorbo A, Panichella S, Visaggio CA, Di~Penta M, Canfora G, Gall HC (2015) Development emails content analyzer: Intention mining in developer discussions (t). In: 2015 30th IEEE/ACM International Conference on Automated Software Engineering (ASE), IEEE, pp 12--23

\bibitem[{Gagniuc(2017)}]{gagniuc2017markov}
Gagniuc PA (2017) Markov chains: from theory to implementation and experimentation. John Wiley \& Sons

\bibitem[{G{\'o}mez et~al.(2013)G{\'o}mez, Cleary, and Singer}]{gomez2013study}
G{\'o}mez C, Cleary B, Singer L (2013) A study of innovation diffusion through link sharing on stack overflow. In: 2013 10th working conference on mining software repositories (MSR), IEEE, pp 81--84

\bibitem[{Guo et~al.(2024)Guo, Cao, Xie, Liu, Li, Chen, and Peng}]{guo2024exploring}
Guo Q, Cao J, Xie X, Liu S, Li X, Chen B, Peng X (2024) Exploring the potential of chatgpt in automated code refinement: An empirical study. In: Proceedings of the 46th IEEE/ACM International Conference on Software Engineering, pp 1--13

\bibitem[{Hata et~al.(2019)Hata, Treude, Kula, and Ishio}]{hata20199}
Hata H, Treude C, Kula RG, Ishio T (2019) 9.6 million links in source code comments: Purpose, evolution, and decay. In: 2019 IEEE/ACM 41st International Conference on Software Engineering (ICSE), IEEE, pp 1211--1221

\bibitem[{Hata et~al.(2022)Hata, Novielli, Baltes, Kula, and Treude}]{hata2022github}
Hata H, Novielli N, Baltes S, Kula RG, Treude C (2022) Github discussions: An exploratory study of early adoption. Empirical Software Engineering 27:1--32

\bibitem[{Hou et~al.(2023)Hou, Zhao, Liu, Yang, Wang, Li, Luo, Lo, Grundy, and Wang}]{hou2023large}
Hou X, Zhao Y, Liu Y, Yang Z, Wang K, Li L, Luo X, Lo D, Grundy J, Wang H (2023) Large language models for software engineering: A systematic literature review. arXiv preprint arXiv:230810620

\bibitem[{Huang et~al.(2018)Huang, Xia, Lo, and Murphy}]{huang2018automating}
Huang Q, Xia X, Lo D, Murphy GC (2018) Automating intention mining. IEEE Transactions on Software Engineering 46(10):1098--1119

\bibitem[{Jiang et~al.(2023)Jiang, Liu, Lutellier, and Tan}]{jiang2023impact}
Jiang N, Liu K, Lutellier T, Tan L (2023) Impact of code language models on automated program repair. In: Proceedings of the 45th International Conference on Software Engineering, IEEE Press, ICSE '23, p 1430–1442, \doi{10.1109/ICSE48619.2023.00125}, \urlprefix\url{https://doi.org/10.1109/ICSE48619.2023.00125}

\bibitem[{Landis and Koch(1977)}]{landis1977measurement}
Landis JR, Koch GG (1977) The measurement of observer agreement for categorical data. biometrics pp 159--174

\bibitem[{Li et~al.(2018)Li, Ren, Li, Zou, and Jiang}]{li2018issue}
Li L, Ren Z, Li X, Zou W, Jiang H (2018) How are issue units linked? empirical study on the linking behavior in github. In: 2018 25th Asia-Pacific Software Engineering Conference (APSEC), IEEE, pp 386--395

\bibitem[{Liang et~al.(2024)Liang, Yang, and Myers}]{liang2024large}
Liang JT, Yang C, Myers BA (2024) A large-scale survey on the usability of ai programming assistants: Successes and challenges. In: Proceedings of the 46th IEEE/ACM International Conference on Software Engineering, pp 1--13

\bibitem[{Liu et~al.(2021)Liu, Xia, Lo, Zhang, Zou, Hassan, and Li}]{liu2021broken}
Liu J, Xia X, Lo D, Zhang H, Zou Y, Hassan AE, Li S (2021) Broken external links on stack overflow. IEEE Transactions on Software Engineering 48(9):3242--3267

\bibitem[{Liu et~al.(2022)Liu, Zhang, Xia, Lo, Zou, Hassan, and Li}]{liu2022exploratory}
Liu J, Zhang H, Xia X, Lo D, Zou Y, Hassan AE, Li S (2022) An exploratory study on the repeatedly shared external links on stack overflow. Empirical Software Engineering 27:1--32

\bibitem[{Lu et~al.(2023)Lu, Yu, Li, Yang, and Zuo}]{lu2023llama}
Lu J, Yu L, Li X, Yang L, Zuo C (2023) Llama-reviewer: Advancing code review automation with large language models through parameter-efficient fine-tuning. In: 2023 IEEE 34th International Symposium on Software Reliability Engineering (ISSRE), IEEE, pp 647--658

\bibitem[{Mozannar et~al.(2022)Mozannar, Bansal, Fourney, and Horvitz}]{mozannar2022reading}
Mozannar H, Bansal G, Fourney A, Horvitz E (2022) Reading between the lines: Modeling user behavior and costs in ai-assisted programming. arXiv preprint arXiv:221014306

\bibitem[{Nijkamp et~al.(2022)Nijkamp, Pang, Hayashi, Tu, Wang, Zhou, Savarese, and Xiong}]{nijkamp2022codegen}
Nijkamp E, Pang B, Hayashi H, Tu L, Wang H, Zhou Y, Savarese S, Xiong C (2022) Codegen: An open large language model for code with multi-turn program synthesis. In: The Eleventh International Conference on Learning Representations

\bibitem[{Nurwidyantoro et~al.(2022)Nurwidyantoro, Shahin, Chaudron, Hussain, Shams, Perera, Oliver, and Whittle}]{nurwidyantoro2022human}
Nurwidyantoro A, Shahin M, Chaudron MR, Hussain W, Shams R, Perera H, Oliver G, Whittle J (2022) Human values in software development artefacts: A case study on issue discussions in three android applications. Information and Software Technology 141:106731

\bibitem[{{OpenAI}(2024{\natexlab{a}})}]{sharedlinkfaq}
{OpenAI} (2024{\natexlab{a}}) {ChatGTP Shared Links FAQ}. \urlprefix\url{https://help.openai.com/en/articles/7925741-chatgpt-shared-links-faq}, accessed: 2024-01-23

\bibitem[{{OpenAI}(2024{\natexlab{b}})}]{openaihelp}
{OpenAI} (2024{\natexlab{b}}) {Create a Shared Link}. \urlprefix\url{https://help.openai.com/en/articles/7943611-create-a-shared-link}, accessed: 2024-01-23

\bibitem[{Qu et~al.(2019)Qu, Yang, Croft, Zhang, Trippas, and Qiu}]{qu2019user}
Qu C, Yang L, Croft WB, Zhang Y, Trippas JR, Qiu M (2019) User intent prediction in information-seeking conversations. In: Proceedings of the 2019 Conference on Human Information Interaction and Retrieval, pp 25--33

\bibitem[{Rosen and Shihab(2016)}]{rosen2016mobile}
Rosen C, Shihab E (2016) What are mobile developers asking about? a large scale study using stack overflow. Empirical Software Engineering 21:1192--1223

\bibitem[{Ross et~al.(2023)Ross, Martinez, Houde, Muller, and Weisz}]{ross2023programmer}
Ross SI, Martinez F, Houde S, Muller M, Weisz JD (2023) The programmer’s assistant: Conversational interaction with a large language model for software development. In: Proceedings of the 28th International Conference on Intelligent User Interfaces, pp 491--514

\bibitem[{Shi et~al.(2021)Shi, Chen, Yang, Jiang, Jiang, Niu, and Wang}]{shi2021first}
Shi L, Chen X, Yang Y, Jiang H, Jiang Z, Niu N, Wang Q (2021) A first look at developers’ live chat on gitter. In: Proceedings of the 29th ACM Joint Meeting on European Software Engineering Conference and Symposium on the Foundations of Software Engineering, pp 391--403

\bibitem[{Siddiq et~al.(2023)Siddiq, Santos, Tanvir, Ulfat, Rifat, and Lopes}]{siddiq2023exploring}
Siddiq ML, Santos J, Tanvir RH, Ulfat N, Rifat FA, Lopes VC (2023) Exploring the effectiveness of large language models in generating unit tests. arXiv preprint arXiv:230500418

\bibitem[{Treude et~al.(2011)Treude, Barzilay, and Storey}]{treude2011programmers}
Treude C, Barzilay O, Storey MA (2011) How do programmers ask and answer questions on the web?(nier track). In: Proceedings of the 33rd international conference on software engineering, pp 804--807

\bibitem[{Vaithilingam et~al.(2022)Vaithilingam, Zhang, and Glassman}]{vaithilingam2022expectation}
Vaithilingam P, Zhang T, Glassman EL (2022) Expectation vs. experience: Evaluating the usability of code generation tools powered by large language models. In: Chi conference on human factors in computing systems extended abstracts, pp 1--7

\bibitem[{Viviani et~al.(2019)Viviani, Famelis, Xia, Janik-Jones, and Murphy}]{viviani2019locating}
Viviani G, Famelis M, Xia X, Janik-Jones C, Murphy GC (2019) Locating latent design information in developer discussions: A study on pull requests. IEEE Transactions on Software Engineering 47(7):1402--1413

\bibitem[{Wan et~al.(2019)Wan, Xia, and Hassan}]{wan2019discussed}
Wan Z, Xia X, Hassan AE (2019) What is discussed about blockchain? a case study on the use of balanced lda and the reference architecture of a domain to capture online discussions about blockchain platforms across the stack exchange communities. IEEE Transactions on Software Engineering (01):1--1

\bibitem[{Wang et~al.(2021)Wang, Xiao, Thongtanunam, Kula, and Matsumoto}]{wang2021understanding}
Wang D, Xiao T, Thongtanunam P, Kula RG, Matsumoto K (2021) Understanding shared links and their intentions to meet information needs in modern code review: A case study of the openstack and qt projects. Empirical Software Engineering 26:1--32

\bibitem[{Wang et~al.(2024)Wang, Chen, Yuan, Zhang, Li, Peng, and Ji}]{wang2024executable}
Wang X, Chen Y, Yuan L, Zhang Y, Li Y, Peng H, Ji H (2024) Executable code actions elicit better llm agents. arXiv preprint arXiv:240201030

\bibitem[{Xiao et~al.(2023)Xiao, Baltes, Hata, Treude, Kula, Ishio, and Matsumoto}]{xiao202318}
Xiao T, Baltes S, Hata H, Treude C, Kula RG, Ishio T, Matsumoto K (2023) 18 million links in commit messages: purpose, evolution, and decay. Empirical Software Engineering 28(4):91

\bibitem[{Xiao et~al.(2024)Xiao, Treude, Hata, and Matsumoto}]{devgpt}
Xiao T, Treude C, Hata H, Matsumoto K (2024) Devgpt: Studying developer-chatgpt conversations. In: Proceedings of the International Conference on Mining Software Repositories (MSR 2024)

\bibitem[{Ye et~al.(2017)Ye, Xing, and Kapre}]{ye2017structure}
Ye D, Xing Z, Kapre N (2017) The structure and dynamics of knowledge network in domain-specific q\&a sites: a case study of stack overflow. Empirical Software Engineering 22:375--406

\bibitem[{Zampetti et~al.(2017)Zampetti, Ponzanelli, Bavota, Mocci, Di~Penta, and Lanza}]{zampetti2017developers}
Zampetti F, Ponzanelli L, Bavota G, Mocci A, Di~Penta M, Lanza M (2017) How developers document pull requests with external references. In: 2017 IEEE/ACM 25th International Conference on Program Comprehension (ICPC), IEEE, pp 23--33

\bibitem[{Zhang et~al.(2023)Zhang, Liang, Zhou, Ahmad, and Waseem}]{zhang2023practices}
Zhang B, Liang P, Zhou X, Ahmad A, Waseem M (2023) Practices and challenges of using github copilot: An empirical study. arXiv preprint arXiv:230308733

\bibitem[{Zhang et~al.(2019)Zhang, Gao, Ma, Lyu, and Kim}]{zhang2019empirical}
Zhang T, Gao C, Ma L, Lyu M, Kim M (2019) An empirical study of common challenges in developing deep learning applications. In: 2019 IEEE 30th International Symposium on Software Reliability Engineering (ISSRE), IEEE, pp 104--115

\bibitem[{Zhang et~al.(2018)Zhang, Yu, Wang, Vasilescu, and Filkov}]{zhang2018within}
Zhang Y, Yu Y, Wang H, Vasilescu B, Filkov V (2018) Within-ecosystem issue linking: a large-scale study of rails. In: Proceedings of the 7th international workshop on software mining, pp 12--19

\bibitem[{Ziegler et~al.(2022)Ziegler, Kalliamvakou, Li, Rice, Rifkin, Simister, Sittampalam, and Aftandilian}]{ziegler2022productivity}
Ziegler A, Kalliamvakou E, Li XA, Rice A, Rifkin D, Simister S, Sittampalam G, Aftandilian E (2022) Productivity assessment of neural code completion. In: Proceedings of the 6th ACM SIGPLAN International Symposium on Machine Programming, pp 21--29

\end{thebibliography}

\end{document}